\documentclass[superscriptaddress,10pt,aps,pra,twocolumn,longbibliography]{revtex4-2}
\usepackage{float}
\floatstyle{boxed}
\usepackage{array}
\usepackage{graphicx}
\usepackage{amsbsy}
\usepackage[utf8x]{inputenc}
\usepackage{epstopdf}
\usepackage{amsmath,amssymb,amsfonts,amsthm}
\usepackage{indentfirst}
\usepackage{soul}
\usepackage[T1]{fontenc}
\usepackage[dvipsnames]{xcolor}
\usepackage{url}
\usepackage[colorlinks]{hyperref}
\usepackage{slashbox}
\usepackage{array}

\hypersetup{
    plainpages=true,
    breaklinks=true,
    hypertexnames=false,
    pageanchor=true,
    colorlinks=true,
    linkcolor={blue},
    citecolor={red},
    urlcolor={blue},
    anchorcolor={black}
}

\newcommand{\figref}[1]{\mbox{Fig.~\ref{#1}}}

\renewcommand{\eqref}[1]{\mbox{Eq.~(\ref{#1})}}

\newcommand{\be}{\begin{equation}}
\newcommand{\ee}{\end{equation}}
\newcommand{\bea}{\begin{eqnarray}}
\newcommand{\eea}{\end{eqnarray}}

\newcommand{\rhot}{\hat{\rho}(t)}

\begin{document}

\title{Topological state permutations in time-modulated non-Hermitian multiqubit systems with suppressed non-adiabatic transitions}

\author{Ievgen I. Arkhipov}
\email{ievgen.arkhipov@upol.cz}
\affiliation{Joint Laboratory of
Optics of Palack\'y University and Institute of Physics of CAS,
Faculty of Science, Palack\'y University, 17. listopadu 12, 771 46
Olomouc, Czech Republic}

\author{Philippe Lewalle}
\affiliation{Department of Chemistry, University of California, Berkeley, CA 94720, USA  and \\ Berkeley Center for Quantum Information and Computation, Berkeley, CA 94720, USA}

\author{Franco Nori}
\affiliation{Quantum Information Physics
Theory Research Team, Quantum Computing Center, RIKEN, Wakoshi, Saitama, 351-0198, Japan} \affiliation{Physics Department, The University of Michigan, Ann Arbor, Michigan 48109-1040, USA}

\author{\c{S}ahin K. \"Ozdemir}
\email{sahin.ozdemir@slu.edu}
\affiliation{Department of Electrical and Computer Engineering, Saint Louis University, St.~Louis, Missouri 63103, USA}

\author{K. Birgitta Whaley}
\affiliation{Department of Chemistry, University of California, Berkeley, CA 94720, USA  and \\ Berkeley Center for Quantum Information and Computation, Berkeley, CA 94720, USA}

\begin{abstract}
Non-Hermitian systems have been at the center of intense research for over a decade, partly due to their nontrivial energy topology formed by intersecting Riemann manifolds with branch points known as exceptional points (EPs). This spectral property can be exploited, e.g., to achieve topologically controlled state permutations that are necessary for implementing a wide class of classical and quantum information protocols. However, the complex-valued spectra of typical non-Hermitian systems lead to instabilities, losses, and breakdown of adiabaticity, which impedes the practical use of EP-induced energy topologies in quantum information protocols based on state permutation symmetries.
Indeed, in a given non-Hermitian multiqubit system, the dynamical winding around EPs always results in a predetermined set of attenuated final eigenstates, due to the interplay of decoherence and non-adiabatic transitions, irrespective of the initial conditions.
In this work, we address this long-standing problem by introducing a model of interacting qubits governed by an effective non-Hermitian Hamiltonian that hosts novel types of EPs while maintaining a completely real energy spectrum, ensuring the absence of losses in the system's dynamics. We demonstrate that such non-Hermitian Hamiltonians enable the realization of genuine, in general, non-Abelian permutation groups in the multiqubit system's eigenspace while dynamically encircling these EPs.
Our findings indicate that, contrary to previous beliefs, non-Hermiticity can be utilized to achieve controlled topological state permutations in time-modulated multiqubit systems, thus paving the way for the advancement and development of novel quantum information protocols in real-world non-Hermitian quantum systems.
\end{abstract}

\date{\today}

\maketitle

\section{Introduction}
The presence of Riemann topology with branch points, called exceptional points (EPs), in the energy spectrum of non-Hermitian systems has attracted considerable interest over the last few decades. The EPs, where non-Hermitian Hamiltonian (NHH) eigenvectors coincide, have been linked to a number of exciting and nontrivial effects and phenomena not found in Hermitian systems~\cite{Ozdemir2019,Ashida2020,El-Ganainy2018,Bergholtz2021}. 

From the early days of research on EPs, it has been suggested that the topology of the intersecting Riemann energy manifolds at EPs can be exploited for the implementation of a symmetric state switch~\cite{Latinne1995,Lefebvre2009,Atabek2011,Heiss2000,Cartarius2007}. 
In other words, by dynamically encircling EPs in the system parameter space, an initial eigenstate, belonging to one of the Riemann energy sheets, adiabatically transitions to higher or lower-lying Riemann energy eigenstates, thereby enabling a symmetric state transfer~\cite{Dembowski2003,Dietz2011,Gao2015,Ding2016,Zhong2018,Ergoktas2022,Guria2024}. 

However, subsequent studies revealed that symmetric state switching via dynamical encirclement of EPs is generally impractical due to the breakdown of adiabaticity in non-Hermitian systems, driven by the presence of 
{a complex valued energy} spectrum~\cite{Nenciu_1992, Uzdin2011, Berry2011, Graefe2013}. In particular, any initial eigenmode of the NHH evolves into the state with the minimal loss {(smallest imaginary energy component)} through a series of non-adiabatic transitions (NATs) during the system dynamics, {imbuing the dynamics with a chiral nature.}  This intriguing chiral mode behavior has also been demonstrated in experiments~\cite{Doppler_2016, Hassan2017, Xu2016, Yoon2018, Zhang2018_encirc, Zhang2019_encirc2, Yu2021, Feng2022, Abbasi2022,Tang2023,Khandelwal2024,Bu2024}, {and various approaches for state control have been explored in the context of such quantum dynamics \cite{Ibanez_2011, Ibanez_2014, Torosov_2013, Funo2020, Chen2021c, Ribiero_2021, Luan_2022, Lewalle_2023, chavva2025topological}.}

{The presence of NATs generically inhibits the possibility of using adiabatic dynamics to realize state--switching dynamics based on following eigensheets in the NHH spectrum.}
{
One possible approach to restoring state transfer is to employ shortcuts to adiabaticity (STA)~\cite{Wu2025, chavva2025topological}, where additional control fields are introduced to replicate the desired outcome of ideal adiabatic evolution.
Alternatively, some progress has been made in restoring symmetric state flips to enable a programmable multi-state switch~\cite{arkhipov2023b}. 
Both of these approaches presume that NATs appear in the system dynamics and are undesirable, and {that one therefore 
seeks} physical mechanisms to mitigate their effects.
}

{Here, we {consider instead the approach of} 
engineering systems in which non-trivial topological features appear in the NHH spectrum, but where NATs never disrupt the eigenstate evolution, thus avoiding the need for additional controls. That is, we {consider here} the realization of genuinely {\it adiabatic} multi-state transfer in a non-Hermitian context.}
In a recent work~\cite{arkhipov2024a} {some of us} {have} demonstrated that {such an approach is possible} 
in two-level systems by carefully choosing state trajectories in the system's parameter space {on a submanifold described by a pseudo-Hermitian Hamiltonian with real eigenvalues}, allowing adiabaticity to be fully restored while dynamically winding around EPs. However, 
while that method can {potentially be} extended to linear optical multimode systems by combining 
{it with the} findings in Ref.~\cite{arkhipov2023b}, it presents a significant challenge when applied to {\it interacting} multiqubit systems. {Moreover, the method in Ref.~\cite{arkhipov2024a} enables topological and adiabatic state transfer only among non-orthogonal eigenstates, {since} the entire system must be initialized in the non-Hermitian phase. This severely limits its applicability and integration with state-of-the-art quantum information protocols.}

{In this work, we address {this} long-standing challenge of {\it adiabatic} multi-state transfer in time-modulated non-Hermitian {multiple qubit} systems by introducing an NHH that describes an interacting multiqubit system and exhibits a purely real spectrum with specific EPs defined on hyperboloid manifolds in the system parameter space. At this special type of EPs, which we refer to as `twisted' EPs, the Hamiltonian eigenvectors 
{coalesce,} although the corresponding eigenvalues diverge, with the sign of the divergence depending on the path taken to approach these EPs.
Remarkably, by dynamically winding around such EPs along the Riemann eigensheets, 
{we show that} one can implement a genuine and controllable symmetric multi-state switching in the system dynamics, characterized by the absence of the NATs.  This facilitates the generation of various permutation groups in the NHH eigenspace, a property that had been previously thought {challenging or impossible} to achieve due to the breakdown of the adiabatic theorem~\cite{Nenciu1980,Uzdin2011,Berry2011}.}
{Moreover, the proposed NHH can also attain a Hermitian phase, where its eigenstates are orthogonal, and over which the multi-state switching protocol can {also} be performed. This allows to merge and combine existing quantum information protocols {that have been developed} 
for Hermitian systems, to non-Hermitian multi-qubit systems as well.}

The ability to generate and control permutation groups in the NHH eigenspace holds promise for  
{new opportunities in} quantum information and communication. Potential applications include the production of diverse entangled states~\cite{Horodecki2009}, the implementation of optimal quantum gates~\cite{Barenco1995}, the development of quantum error-correcting codes~\cite{Kada2008, Buhrman2024}, performing holonomic quantum computations~\cite{Zhang_2023,Zhang2018_hol,Chen2021_hol}, and advancements in quantum sensing~\cite{Degen2017}, among others.

As an illustrative example, we study {here} an NHH describing a three-qubit system and investigate the rich structure of topological eigenstate permutations that arise while dynamically orbiting its {associated} EPs. {{These eigenstate permutations are topological because 
they} do not depend on the shape of the winding loops around the EPs, 
{and} are {determined solely} by the intrinsic Riemann topology in the vicinity of the EPs.} Our analysis reveals that the symmetric state switching protocol implemented 
{in this work} can generate a nontrivial non-Abelian permutation group comprising 576 elements. {We also discuss the robustness of the protocol 
and its potential realization in various existing experimental setups.}
Our work therefore opens avenues for advancing and developing novel quantum information protocols {using non-Hermitian quantum systems}, {in which} generating and steering permutations {of quantum states} plays a pivotal role.

\section{Model}

We start by introducing a tight-binding non-Hermitian Hamiltonian $\hat H\neq \hat H^{\dagger}$ describing $N$ qubits represented as spins with $ZZ$ coupling, 
which are subject to a transverse, in general non-uniform, effective {complex-valued} magnetic field $\vec{\bold B}_{||}^k$:
\begin{equation}\label{H}
    \hat H=\sum\limits_k^N\vec{{\boldsymbol\sigma}}^k\vec{\bold B}_{||}^k+\sum\limits_{k<l}J_{kl}\hat\sigma_z^k\hat\sigma_z^l.
\end{equation}
Here, the spin vector $\vec{{\boldsymbol\sigma}}=[\hat\sigma_x,\hat\sigma_y,0]^T$, with the symbol $T$ denoting the transpose operation, $\hat\sigma_{x,y,z}$ are the Pauli matrices, and the {transverse} magnetic field vector {has $x, y$ components and} is written as
\begin{eqnarray}\label{B}
    \vec{\bold B}_{||}^k=f_k\Big[\alpha\sin\phi,\alpha\cos\phi,0\Big]^T,
\end{eqnarray}
where
\begin{eqnarray}\label{alpha}
    \alpha=\frac{y\sin\left({\rm Re}[\phi]\right)}{\sin\left({\rm Im}[\phi]\right)}, \quad \phi=\arctan\left[(x+iy)^{-1}\right],
\end{eqnarray}
with $f_k,x,y\in{\mathbb R}$. 
{For fixed $f_k$, the system parameter space is defined by a tuple $(x,y,\textbf{J})$, with $\textbf{J}$ being the coupling matrix $\textbf{J}\equiv J_{kl}$.} 

The equations~(\ref{B}) and (\ref{alpha}) determine the embedding map $(x,y)\to (B_{\rm x}^r,B_{\rm x}^i,B_{\rm y}^r,B_{\rm y}^i)$, where the superscripts in 
{the four $B$ components} denote the real and imaginary parts of the magnetic field {in the $(\rm x)$ and $(\rm y)$ directions.}
The components of the local magnetic field thus describe a hyperboloid surface with a varied curvature $\alpha$ in ${\mathbb R}^4$~\cite{arkhipov2024a}, namely
\begin{eqnarray}
    (B_{\rm x}^r)^2+(B_{\rm y}^r)^2-(B_{\rm x}^i)^2-(B_{\rm y}^i)^2\equiv\alpha^2.
\end{eqnarray}
This parametrization of the magnetic field, together with the $ZZ$ qubit interactions, ensures that the NHH in \eqref{H} is pseudo-Hermitian and {that} this symmetry is exact in the 
{entire} parameter space, meaning the spectrum of $\hat H$ is everywhere real. Indeed, the condition for the pseudo-Hermiticity of the NHH reads as~\cite{MOSTAFAZADEH_2010} 
\begin{eqnarray}
    \eta\hat H\eta^{-1}=\hat H^{\dagger},
\end{eqnarray}
where $\eta$ is some Hermitian operator. Here, the operator $\eta$ can be chosen as a metric operator of the NHH describing non-interacting qubits, i.e., when $\textbf{J}=\boldsymbol{0}$ (see 
Appendix~\ref{Metric}).

The NHH {of} \eqref{H} with {its complex}{-valued} 
{magnetic} fields can be implemented in various experimental platforms, ranging from solid-state systems  to photonic setups, {by making use of}
either the Naimark dilation method~\cite{Gunter2008} or photon postselection techniques~\cite{Minganti2022}.
We elaborate on {possible experimental realizations} in more detail in Sec.~\ref{Discussion}.

\section{Eigenspectrum of the non-Hermitian Hamiltonian and `twisted' exceptional points }\label{sec:eigen}
{We now} discuss the spectral properties of the NHH in \eqref{H}, and the potential {offered by this system} {to realize a genuine adiabatic permutation subgroup in non-Hermitian multiqubit systems with EPs.}

\subsection{Non-interacting qubits}\label{nonint}
\begin{figure}[t!]
    \includegraphics[width=0.39\textwidth]{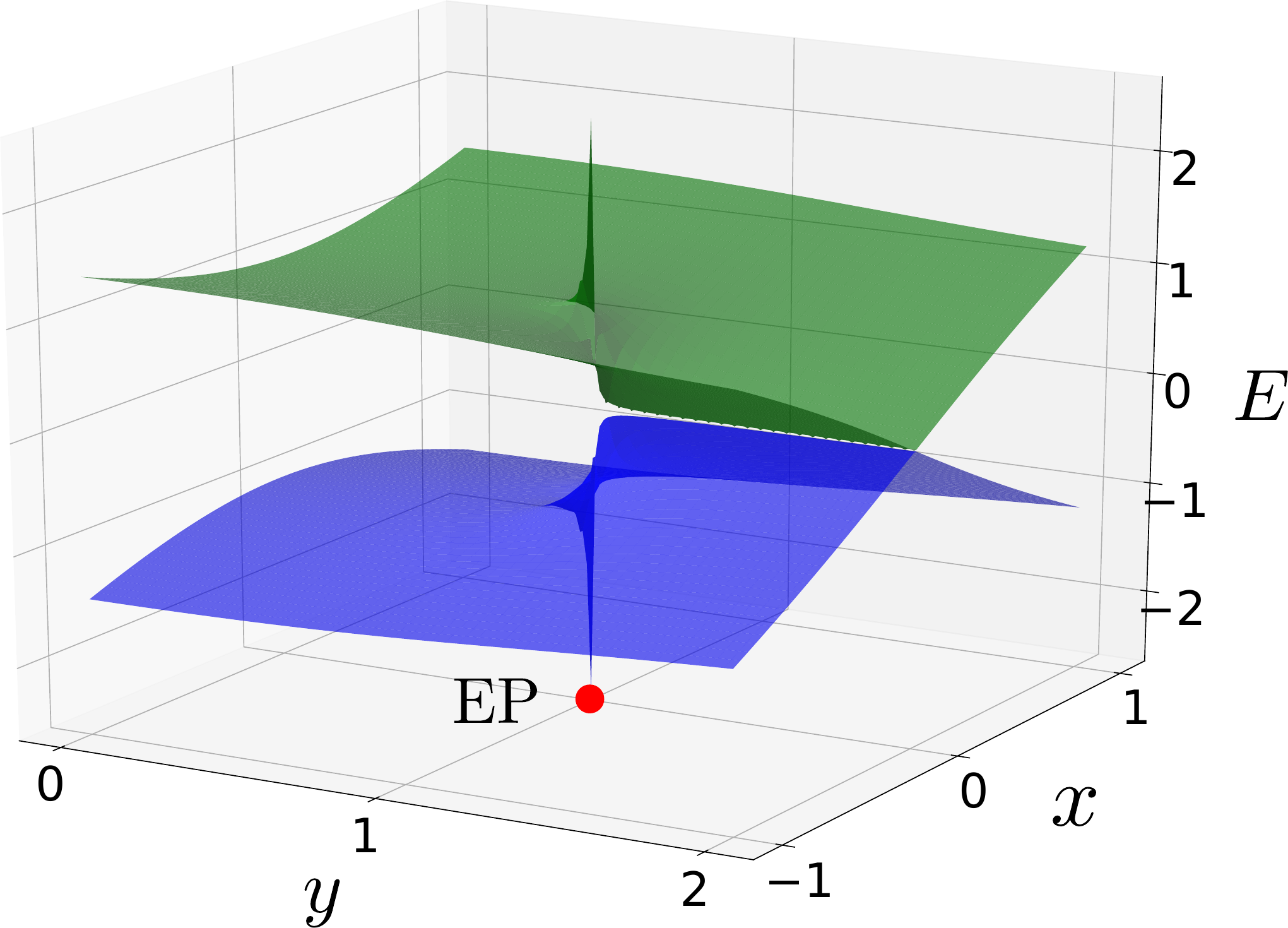}
    \caption{Energy spectrum of a single qubit, described by the NHH in \eqref{H},  as a function of the applied magnetic field $\vec{\bold B}_{||}^i(x,y)$, according to Eqs.~(\ref{E0}) and (\ref{alpha}). {The $x,y$ dependence of the energies reflects the $x,y$ dependence of the parameter $\alpha$.} The purely real-valued spectrum consists of two intertwined Riemann sheets with an  EP located at $(x=0,y=1)$. Moreover, at this EP the spectrum of the NHH diverges, according to \eqref{lim} (see the main text for details).}
    \label{fig1}
\end{figure}

We begin our analysis of the spectral properties of the NHH in \eqref{H} with the simplest case, when $\textbf{J}\equiv\boldsymbol{0}$, and $f_k=1$, for $\forall k$. In {this} case the spectrum of $N$ non-interacting qubits is analogous to that of an $N$-dimensional hypercube~\cite{arkhipov2023c,arkhipov2024b}, namely
\begin{eqnarray}\label{E0}
    E_k\equiv-N\alpha,-(N-2)\alpha,\dots,(N-2)\alpha,N\alpha,
\end{eqnarray}
which is generated by the spectrum of a single qubit $E^{\rm q}_{1,2}=\pm\alpha$. 
{This single qubit energy spectrum is plotted} in \figref{fig1}.
In what follows, we assume that the energy levels are always arranged in ascending order, i.e.,  
\begin{eqnarray}\label{order}
    E_1\leq E_2\leq \dots \leq E_{2N}.
\end{eqnarray}
The right eigenvector space of the non-interacting qubits' NHH is spanned by tensor products of the eigenvectors of each individual qubit that read~\cite{arkhipov2023c,arkhipov2024a}
\begin{eqnarray}\label{psi}
   |\psi_1\rangle\equiv \hat R\begin{pmatrix}
        -\sin\dfrac{\phi}{2} \\
        \cos\dfrac{\phi}{2}
    \end{pmatrix}, \quad
    |\psi_2\rangle\equiv \hat R\begin{pmatrix}
        \cos\dfrac{\phi}{2} \\
        \sin\dfrac{\phi}{2}
    \end{pmatrix},
\end{eqnarray}
where $\hat R=\exp\left(-i\pi\sigma_x/4\right)$ is a rotation operator. 
Because the pseudo-Hermiticity of the parameterized NHH does not 
{allow} spontaneous symmetry breaking, the left eigenvectors are simply the complex conjugates of the right eigenvectors~\cite{arkhipov2024a}.

We now proceed with a more detailed examination of the spectral properties of a single qubit, 
{since} its properties are directly reflected in \eqref{E0}.
As can be seen from \figref{fig1}, the qubit spectrum is characterized by the presence of a singular point at $(x=0,y=1)$. Indeed, according to \eqref{alpha} 
{we have}
\begin{eqnarray}\label{lim}
    \lim\limits_{x\to0,y\to1}\alpha=\pm\infty,
\end{eqnarray} 
where the sign at infinity depends on the direction of the limit approach to the singularity (see 
\figref{fig1}). 
Furthermore, one finds the limit for the inner product of the normalized eigenvectors  at this point 
is 
\begin{eqnarray}\label{lim2}
    \lim\limits_{x\to0,y\to1}|\langle\psi_2|\psi_1\rangle|=1.
\end{eqnarray} 
Equation~(\ref{lim2}) implies that the qubit eigenvectors coalesce at this singular point. 
In other words, the singularity behaves like an EP, {although} it is not an EP in the usual sense that is encountered in {non-Hermitian} systems with finite spectra. 
This seeming difficulty can be handled using mathematical tools borrowed from the projective space framework~\cite{NakaharaBook}. 
Specifically, this singularity can be cast into an ordinary EP in some extended NHH, where the eigenvalues converge at the EP (see Appendix~\ref{AB} for details).

Nevertheless, for our purposes here, we can safely ignore the infinite nature of the spectrum at such EPs, because we are solely interested in the loop trajectories around the EPs, along which both the energy spectrum and the 
parameters controlling the system ($x, y, {\textbf{J}}$), or equivalently, the complex magnetic fields and the qubit-qubit interactions 
remain finite.
These EPs may therefore be termed as 'twisted EPs,' due to the nontrivial wrapping of the spectrum at these points that leads to the divergent behavior of the NHH eigenvalues. For simplicity, however, {in the remaining technical sections of this paper we shall} continue to refer to such singularities as EPs. 

When dynamically encircling the EP in \figref{fig1} in the parameter space $(x,y)$, the eigenstates of a qubit can be interchanged adiabatically without invoking any non-adiabatic transitions~\cite{arkhipov2024a}. This stems from the fact that the eigenvalues are everywhere real. {It is important to appreciate however,} that while a winding trajectory around the EP in the space $(x,y)$ can have the form of simple loops in a plane, the actual orbits in the vector space (${\mathbb C}^2$ or ${\mathbb R}^4$) of the magnetic field $ \vec{\bold B}_{||}$ are the loops on a curved hyperboloid.

We note that the spectrum described above {differs drastically} from that studied in Ref.~\cite{arkhipov2024a}, where the qubit eigenvalues have a well-defined limit in \eqref{lim}. The key distinction lies in the fact that, unlike in Ref.~\cite{arkhipov2024a}, 
the Riemann energy sheets are {now} well separated at the {special} point $(x=0,y=0)$ where the NHH becomes Hermitian (see \figref{fig1}). This is a very useful property, {since} it can {\it allow {one to perform} state transfer protocols over multiqubit states} which are often generated and exploited in closed systems. 
{This key property of the NHH of \eqref{H} for non-interacting qubits} thus enables the combination and extension of existing quantum information protocols, originally devised for Hermitian systems, to {\it non}-Hermitian {qubit} systems as well. 
{In the next subsection we extend this to the case of interacting qubits.}

\begin{figure}[t!]
    \includegraphics[width=0.49\textwidth]{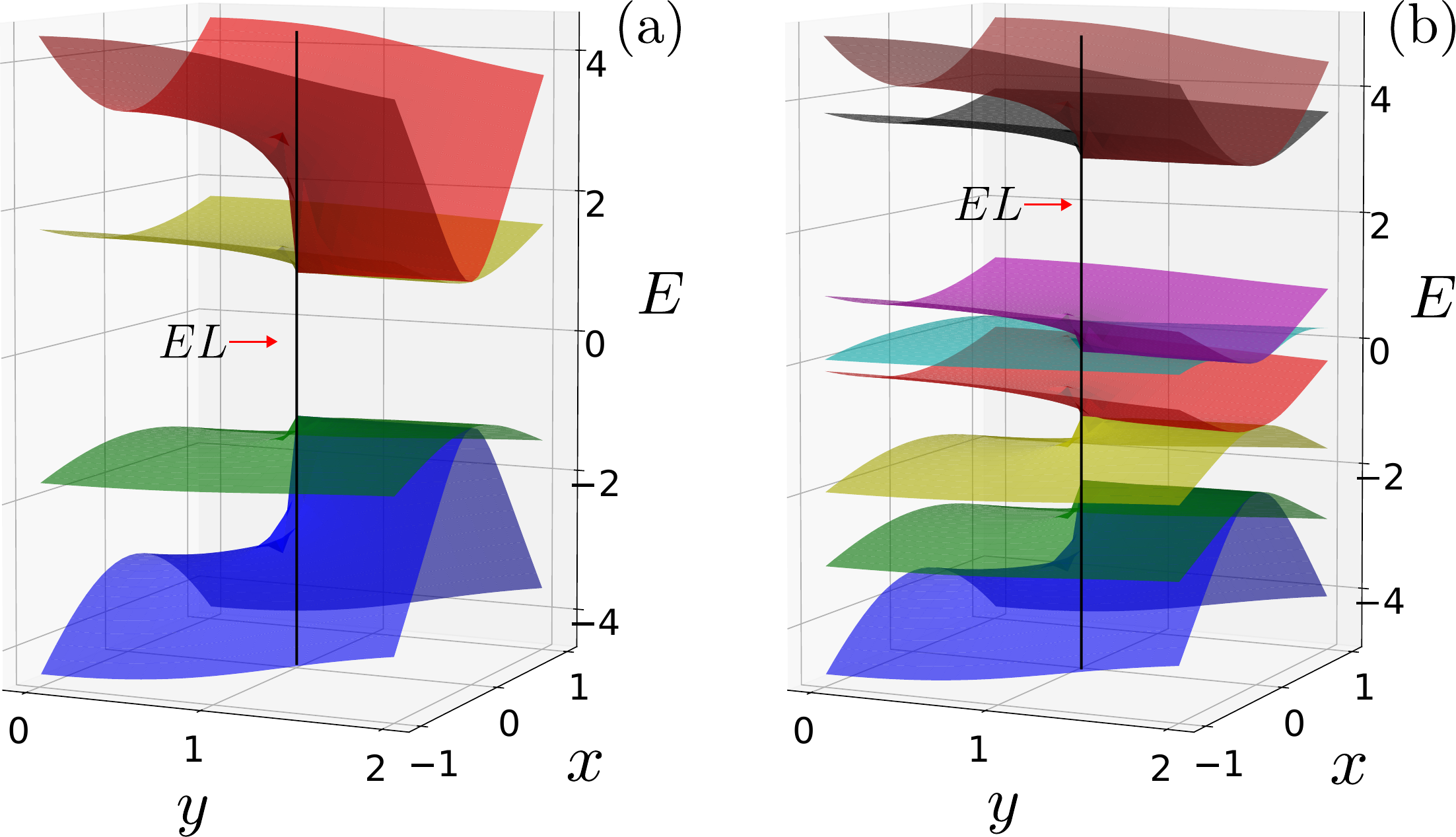}
    \caption{Eigenspectrum of the NHH $\hat H$ in \eqref{H} as a function of the magnetic field $\vec{\bold B}_{||}^i(x,y)$ for (a) two interacting qubits and (b)  three interacting qubits. The system parameters are (a) $f_2=2f_1=2$, $J_{12}=1$; and (b) $f_{1,2,3}=1$, $J_{12}=0.5$, $J_{13}=1$; $J_{23}=1.5$. The spectrum consists of a stack of Riemann sheet pairs, whose number grows with the number of interacting qubits in the system. The set of EPs, constituting the exceptional line (EL), is located at the same point $(x=0,y=1)$ as in the case of a single non-interacting qubit in \figref{fig1}. {We note that the EPs are characterized by the same divergence behavior as in \figref{fig1}, in accordance with \eqref{lim}. As noted in the text, this does not affect the dynamics of trajectories around the EPs so, for the sake of clarity, we omit explicit depiction of the divergence here and in the figures that follow.}}
    \label{fig2}
\end{figure}

\subsection{Interacting qubits}
When the qubit interactions  in \eqref{H} are switched on, the NHH spectrum takes the form of a stack of Riemann sheet pairs, {determined by the function $\alpha(x,y)$,}  with the spacing between sheet pairs determined by a function of the $ZZ$ coupling strengths $J_{kl}$, as
illustrated in \figref{fig2}.
Hence, for interacting qubits, the EP of a single qubit extends to an exceptional line (EL), as depicted in \figref{fig2}. This EL is orthogonal to the $(x, y)$ plane and intersects it at the point $(x = 0, y = 1)$.

To determine whether a non-Hermitian multiqubit system enables a multistate switch, one can examine a topological invariant associated with EPs {that is} known as chirality~\cite{Ding2016}. {Specifically}, when the chirality is non-zero, the eigenstates can be swapped while encircling an EP.
In contrast, when the chirality is zero, winding around the EP does not induce a state flip; instead, each eigenstate returns to itself after completing a full encircling cycle~\cite{Ding2016}. 

A non-zero chirality can be inferred from the behavior of the spectrum near the Riemann branch cuts where $\alpha=0$, 
which corresponds to the region $x=0, y \in [1, \infty)$ in the $(x, y)$ plane (see Figs.~\ref{fig1} and \ref{fig2}). 
Indeed, at the branch cut, the function $\alpha$ changes sign. This means
that if a decomposition of $E_k$ in this region contains odd powers of $\alpha$, 
this will be an indication that the levels interchange when crossing a branch cut.

\section{Implementing a state permutation group in a non-hermitian time-modulated three-qubit system}\label{IV}
\begin{figure}[t!]
    \includegraphics[width=0.49\textwidth]{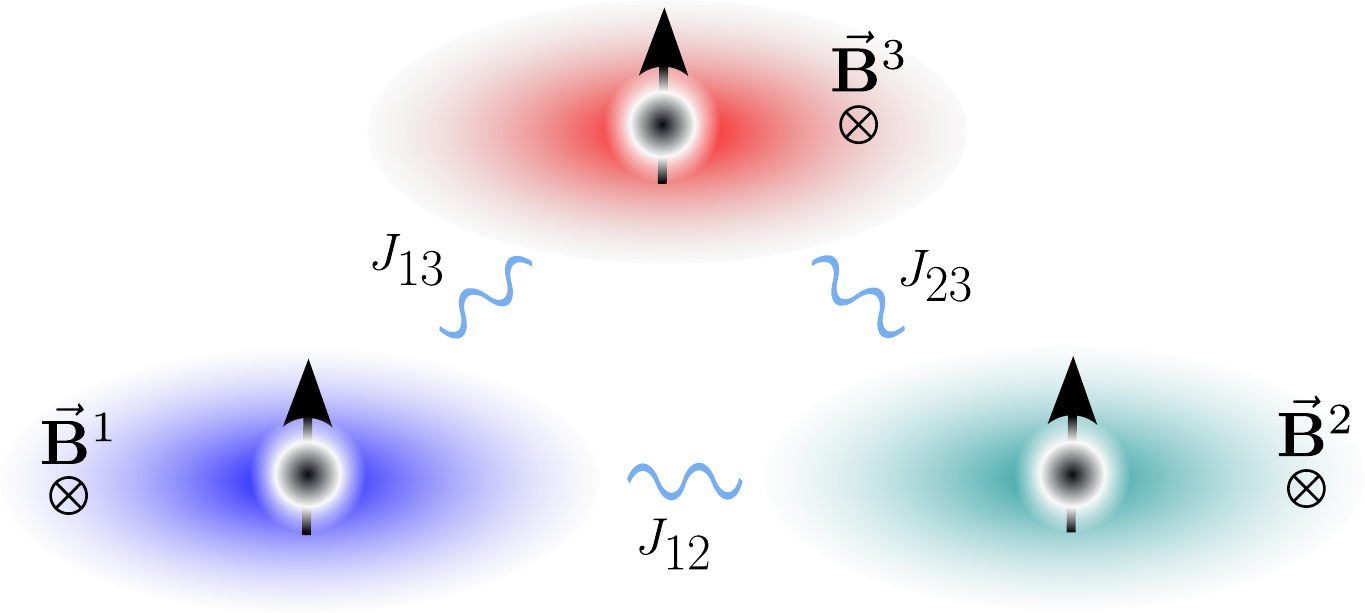}
    \caption{Schematic representation of a three-qubit spin system with pairwise $ZZ$ interactions, subject to a {complex} transverse magnetic field (simplified here as a vector perpendicular to the plane). The system is governed by the effective NHH {of} \eqref{H}. {In general,} the magnetic field strength acting on each spin qubit 
    {is different}, as indicated by the differently colored clouds surrounding the qubits. By appropriately time-modulating {the magnetic fields $\vec{\bold B}^k$ and the qubit interactions $J_{kl}$ of Eq.~(\ref{H}) according to \eqref{param}}, 
    one can realize a nontrivial permutation group acting on the system's NHH eigenspace.}
    \label{fig3}
\end{figure}
We now proceed to describe the implementation of a {\it nontrivial state permutation group} in the eigenspace of a time-modulated three-qubit system as an illustrative example. {A schematic of the protocol is shown in \figref{fig3}}.

\subsection{Description of the state transfer protocol}
\begin{figure*}[t!]
    \includegraphics[width=0.99\textwidth]{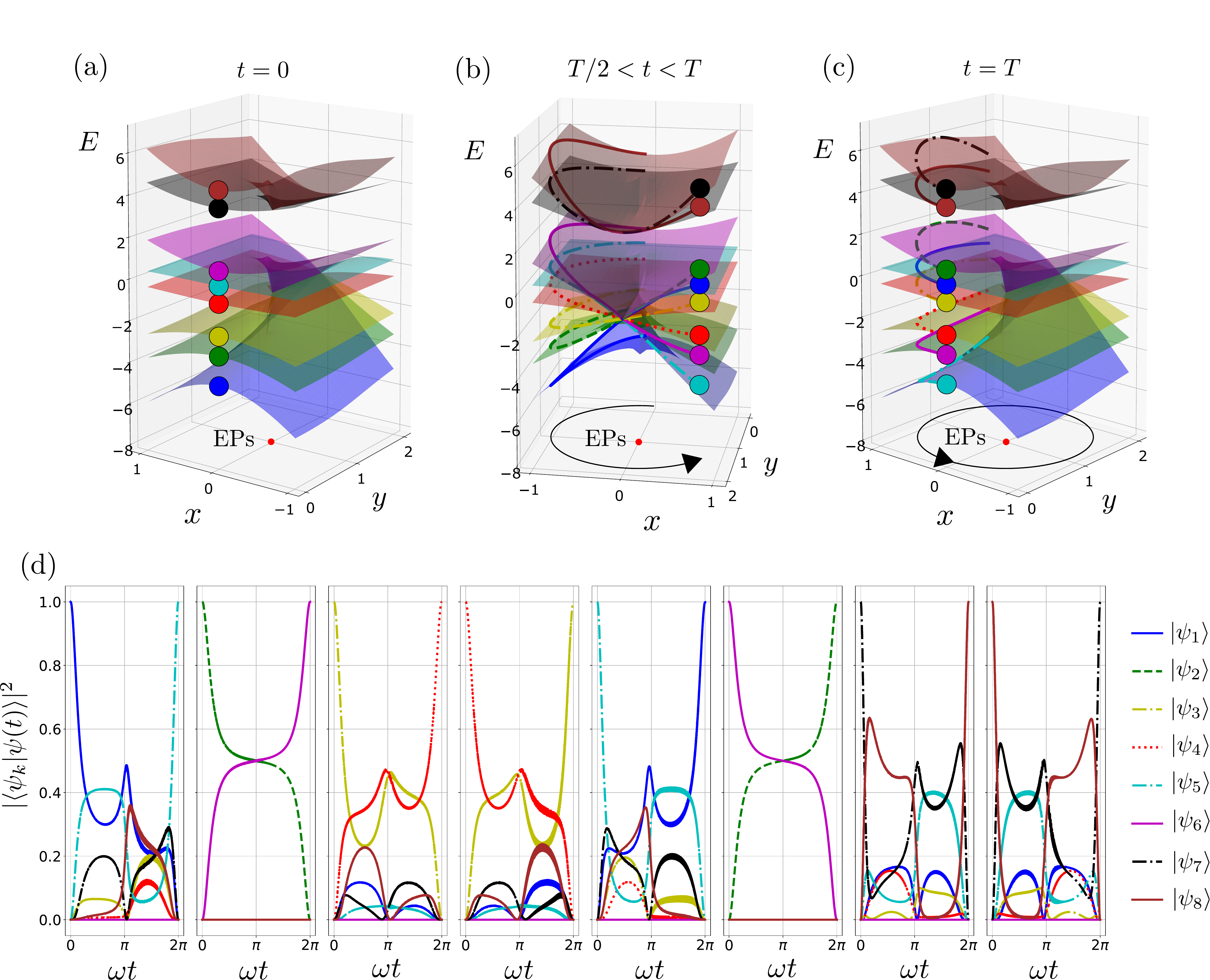}
    \caption{
    Dynamically {encircling} around EPs in a three-qubit system governed by the NHH in \eqref{H} by time modulating the magnetic field $\vec{\bold B}_{||}^i(x,y)$, according to \eqref{param}, over a winding period $T$. (a) A schematic representation of the initial eigenstates at $t=0$, depicted as colored spheres, with each color corresponding to a certain Riemann energy manifold.
    (b),(c) The evolution of various initialized states at times $T/2<t<T$ and at $t=T$, respectively. 
    (d) Fidelity $|\langle \psi_k|\psi(t)\rangle|^2$ between the evolving state $|\psi(t)\rangle$ and the initial eigenstate $|\psi_k\rangle$, $k=1,\dots,8$, of the NHH.
System parameters are set as $f_3=2f_1=2f_2=2$, $J_{12}=J_{13}=J_{23}=1$. {The time period is $T=2500$ [arb. units], the semi-minor and semi-major axis of the enclosing ellipse are $r_x=3$, and $r_y=6$, respectively,}  and the initial phase $\phi_0=\pi$.
    }
    \label{fig4}
\end{figure*}
\begin{figure*}[t!]
    \includegraphics[width=0.99\textwidth]{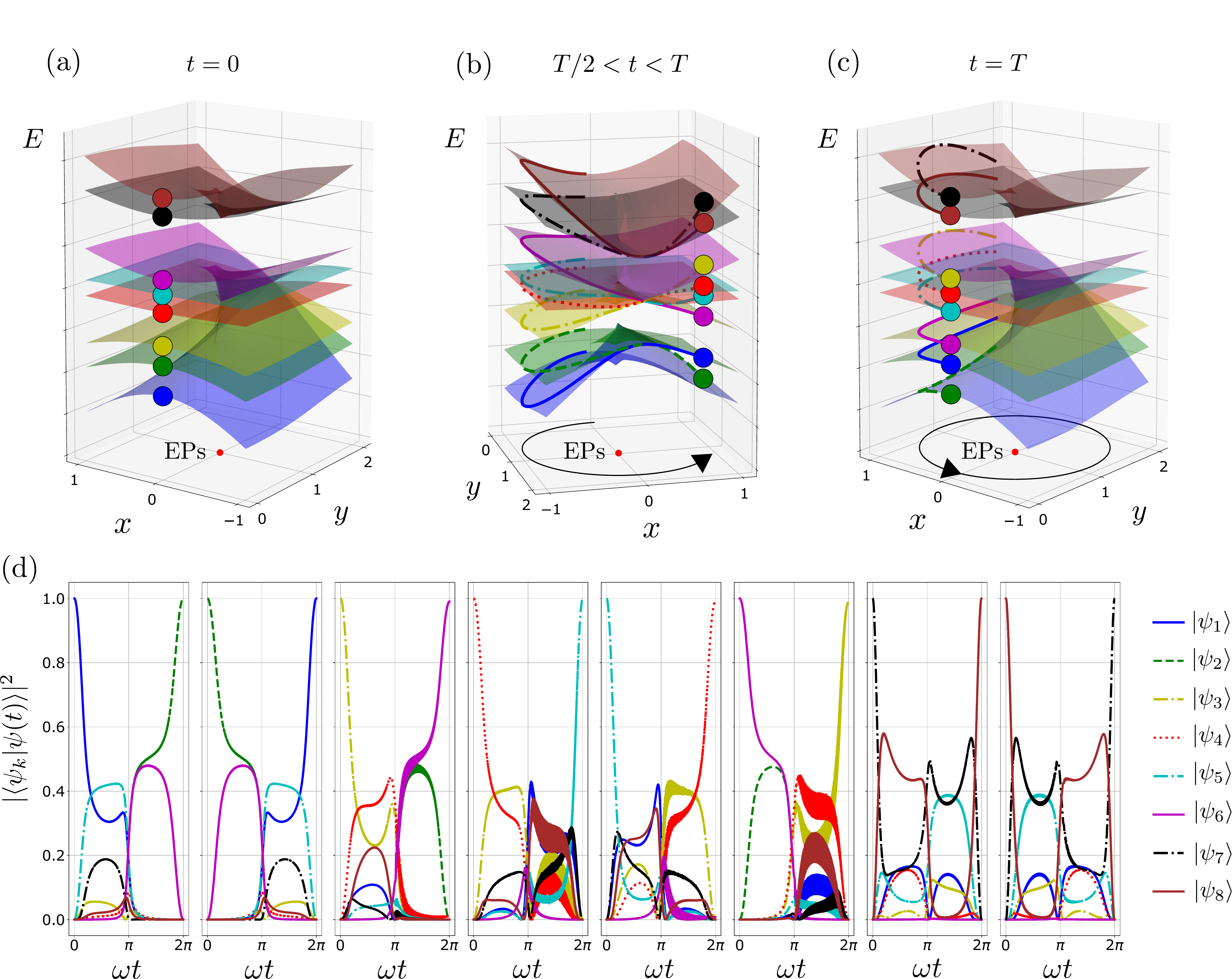}
    \caption{
    Dynamically {spiraling} EPs in a three-qubit system governed by the NHH in \eqref{H} by time modulating both the magnetic field $\vec{\bold B}_{||}^i(x,y)$ and qubit coupling $J_{13}$, according to \eqref{param}, over a winding period $T$. 
    (a) A schematic representation of the initial eigenstates at $t=0$, depicted as colored spheres, with each color corresponding to a certain Riemann energy manifold. 
    (b),(c) The evolution of various initialized states at times $T/2<t<T$ and at $t=T$, respectively. 
    The Riemann energy sheet configuration shown in panel (b) corresponds to $t=T/2$ {and is distinct from that in panel (a), showing distortions and shifts} as a result of the additional modulation of the qubit coupling $J_{13}(t)$. This enables more couplings between different Riemann energy sheets {than for the situation} in \figref{fig3} with static interactions. 
    (d) Fidelity $|\langle \psi_k|\psi(t)\rangle|^2$ between the evolving state $|\psi(t)\rangle$ and the initial eigenstate $|\psi_k\rangle$, ($k=1,\dots,8$), of the given three-qubit NHH. The rest of the system parameters are the same as in \figref{fig3}. We note that the same result is obtained when modulating the qubit coupling $J_{23}$ in \eqref{param}. 
    }
    \label{fig5}
\end{figure*}

In \figref{fig2}(b) we present a typical energy spectrum for {three interacting} qubits, governed by the NHH {of} \eqref{H}.
As can be seen from the figure,  
winding around EPs {that is confined within} the $(x,y)$ plane can only result in the state transfer between eigenstates belonging to individual Riemann pairs. In order to achieve {state flips} between different Riemann pairs as well, one has to ensure that these sheets can be `transported' along the $z$-axis in \figref{fig2}(b). {Our approach is motivated by} the {programmable switch} studied in Ref.~\cite{arkhipov2023b}, where individual pairs of Riemann manifolds of a bosonic multimode spectrum were {similarly} manipulated.
{In the current work,} such control can be achieved by further modulating the interaction terms $J_{kl}$ in the NHH, {Eq.~(\ref{H})}. 
Specifically, we {shall} require the system parameters to be time-modulated  as follows:
\begin{eqnarray}\label{param}
    x(t) &=& r_x\sin(\omega t + \phi_0), \nonumber \\
    y(t) &=& r_y[1 + \cos(\omega t + \phi_0)], \nonumber \\
    J_{kl}(t) &=& J_{kl}\cos^2({\omega t}/{2}), 
\end{eqnarray}
{where $r_{x,y}$ define the semi-minor and/or semi-major axis of the enclosing ellipse in the $(x,y)$ plane,} $\omega =2\pi/T$ is an angular frequency determined by a winding period $T$, and $\phi_0$ is the initial phase. {Clearly, when $r_x=r_y$, the ellipse loop reduces to {a} circle. In this work, however, we do not restrict ourselves exclusively to circular orbits. Indeed, the topological nature of the state transfer 
{resulting from} dynamically encircling EPs should not depend on the shape of the encircling loop.} The qubit coupling strengths $J_{kl}$, with $k,l=1,2,3$, and $k<l$, can be modulated individually, i.e., {it is not necessary to modulate them} simultaneously. To track {the} state evolution in such a time-modulated system, we numerically solve the Schr\"odinger equation $i\partial_t|\psi(t)\rangle=H|\psi(t)\rangle$ with the Runge-Kutta method.

Alternatively, one 
{can} describe the trajectory in \eqref{param} as {\it dynamical spiraling} around EPs rather than mere {\it dynamical encirclement}. This is because, in addition to orbiting a circular path in the 2D $(x, y)$ parameter space, the system is simultaneously driven along a perpendicular axis
defined by the {time-dependent couplings $J_{kl}(t)$,} as illustrated in \figref{fig2}.
{This dynamical spiraling around the EPs is thus similar to the protocol described in Ref.~\cite{arkhipov2023b}, except {for} the {key} fact that {in the current situation the dynamics of the system} now remain purely {\it adiabatic}}.

\subsection{{Adiabatic} dynamical encirclement and spiraling around exceptional points in a three-qubit system }

To elucidate the state transfer characterized by the overlap of distinct Riemann pairs during the dynamical encirclement and spiraling of EPs, we {now} give {two} illustrative examples {for the {interacting} three-qubit system} in Figs.~\ref{fig4} and \ref{fig5}, respectively.

{In \figref{fig4}, we assume} that the time modulation is only performed over the parameters $(x,y)$, leaving interactions unchanged $J_{12}=J_{13}=J_{23}=1$, i.e., we 
{implement only the dynamical encirclement of EPs and no spiralling around these.} 

We initialize the systems in one of the NHH eigenstates, as depicted by the colored spheres in \figref{fig4}(a).
The initial eigenstates $\psi_i$ occupy their respective Riemann energy sheets $E_i$, which are ordered according to \eqref{order}, in the Hermitian phase of the non-Hermitian Hamiltonian $\hat{H}$, where the complex magnetic field $\vec{\bold B} = 0$, i.e., $(x=0, y=0)$. In Figs.~~\ref{fig4}(a)--(c), the evolving eigenstates $\psi_i$ are represented as colored spheres: $\psi_1$ (blue), $\psi_2$ (green), $\psi_3$ (yellow), $\psi_4$ (red), $\psi_5$ (cyan), $\psi_6$ (magenta), $\psi_7$ (black), and $\psi_8$ (brown).

After initiating the counterclockwise encircling of the EPs in the $(x,y)$ plane, {at time $t=T/2$ the evolving states cross the corresponding energy branch cuts, resulting in the state transfer shown in panels (b) and (c) of} \figref{fig4}.
Specifically, due to the couplings between energy manifolds $E_1\leftrightarrow E_5$,  $E_2\leftrightarrow E_6$, $E_3\leftrightarrow E_4$, and $E_7\leftrightarrow E_8$ in the region $[x=0,y>1]$, the evolving states transition as follows [see \figref{fig4}(b)]
\begin{eqnarray}
&\psi_1\rightarrow E_5,\ \psi_2\rightarrow E_6,\ \psi_3\rightarrow E_4,\ \psi_4\rightarrow E_3,& \nonumber \\
&\psi_5\rightarrow E_1,\ \psi_6\rightarrow E_2,\ \psi_7\rightarrow E_8,\ \psi_8\rightarrow E_7.&
\end{eqnarray}
This obtained configuration of evolving eigenstates remains unchanged for the rest of the evolution until the end of the winding period $t = T$. The resulting permutation can be symbolically written in a cyclic notation as 
\begin{eqnarray}
    p_1=(\psi_1,\psi_5)(\psi_2,\psi_6)(\psi_3,\psi_4)(\psi_7,\psi_8),
\end{eqnarray}
where the numbers in parentheses account for the cyclic permutations between corresponding eigenvectors. For instance, the notation $(\psi_1,\psi_5)$ means that the pair of eigenstates $\psi_1$ and $\psi_5$ are permuted with each other.
\color{black}

Figure~\ref{fig4}(d) shows the fidelity between the evolving state $\psi(t)$ and {the} initial eigenstates of the NHH at $t=0$ 
when winding around the EPs is realized in the counterclockwise direction. 
{The fidelities show a clear switching behavior between an initial eigenstate at $t=0$ and a different eigenstate at $t=T$, confirming the perfect state transfer between eigenstates.}
{Identical} results are obtained when winding in the opposite (i.e., clockwise) direction, because the non-adiabatic jumps, and associated with them chiral mode switching, 
{have been}
eliminated in the system due to the real spectrum of the NHH.  

It is evident that with the fixed qubit couplings described above, {while we see perfect eigenstate transfer,} the system is {still} significantly limited in the number of realizable eigenstate flips.
To overcome this restriction one {can} additionally modulate the coupling between, e.g., the first and third qubits, i.e., by modulating the parameter $J_{13}(t)$ in \eqref{param}. {This gives rise to a dynamical spiraling around the EP,} {as well as to time-dependent modifications of the Riemannian energy manifolds.}

We demonstrate the outcome of such a dynamical spiraling around EPs in \figref{fig5}.  The set of possible initialized eigenstates in \figref{fig5}(a) are the same as in \figref{fig4}(a). By modulating all three parameters, $[x(t), y(t), J_{13}(t)]$, an alternative combination of state flips can be achieved. 

{{When} the system parameters are modulated according to~\eqref{param} {with the time-dependent coupling $J_{13}(t)$ now included}, the Riemann sheets begin to deform and {also to} shift along the $z$-axis, as shown in 
{Figs. \ref{fig5}(b) and (c).}  
At one half of the winding period, $t = T/2$ [$x = 0$, $y>1$], several energy manifolds become coupled [see \figref{fig5}(b)]: 
\begin{eqnarray}\label{Ep2}
    E_1 \leftrightarrow E_2,\  E_3 \leftrightarrow E_6,\  E_4 \leftrightarrow E_5,\  E_7 \leftrightarrow E_8.
\end{eqnarray}
These couplings between energy sheets induce transitions among the eigenstates, analogous to the scenario shown in \figref{fig4}.
However, due to the {additional} modulation of $J_{13}$, the specific pairings of energy surfaces at $t = T/2$ {that are determined by} 
\eqref{Ep2} differ from those in \figref{fig4}(b). Consequently, via this dynamical spiraling around the EPs, the following permutation is implemented:
\begin{eqnarray}
    p_2=(\psi_1,\psi_2)(\psi_3,\psi_6)(\psi_4,\psi_5)(\psi_7,\psi_8).
\end{eqnarray}
{These state permutations are confirmed by calculations of the state fidelity, depicted in \figref{fig5}(d).}}
We note {furthermore} that the same {permutation} result can also be achieved {by modulating the coupling $J_{23}(t)$ instead of $J_{13}(t)$.}

\subsection{Implementing a nontrivial permutation group in a three-qubit system}\label{Group}
By choosing various combinations of the {time-}modulated qubit couplings $J_{kl}(t)$, 
{together} with {time-modulated parameters $(x(t),y(t))$} in \eqref{param}, six different permutations can be realized in the given three-qubit system. These permutations can be compactly expressed using cyclic notation as follows:
\begin{eqnarray}\label{p}
      p_1&=&(\psi_1,\psi_5)(\psi_2,\psi_6)(\psi_3,\psi_4)(\psi_7,\psi_8), \nonumber \\
      p_2 &=& (\psi_1,\psi_2)(\psi_3,\psi_6)(\psi_4,\psi_5)(\psi_7,\psi_8), \text{ if $(J_{13})$ or $(J_{23})$}, \nonumber \\
      p_3 &=& (\psi_1,\psi_3)(\psi_2,\psi_6)(\psi_4,\psi_8)(\psi_5,\psi_7), \text{ if $(J_{13},J_{23})$}, \nonumber \\
      p_4 &=& (\psi_1,\psi_3)(\psi_2,\psi_6)(\psi_4,\psi_5)(\psi_7,\psi_8), \text{ if $(J_{12})$}, \nonumber \\
      p_5 &=& (\psi_1,\psi_3)(\psi_2,\psi_4)(\psi_5,\psi_7)(\psi_6,\psi_8), \text{ if $(J_{12},J_{13})$}, \nonumber \\
      p_6 &=& (\psi_1,\psi_8)(\psi_2,\psi_6)(\psi_3,\psi_7)(\psi_4,\psi_5), \text{ if $(J_{12},J_{13},J_{23})$}, \nonumber \\ 
\end{eqnarray}
where the parameters in the parenthesis $(J_{kl},\dots,J_{k'l'})$ indicate the qubit couplings that must be modulated to realize the corresponding permutation. We confirm these findings {numerically} in Appendix~\ref{Permutations}, {where we also provide further examples and discussion of the time-dependent distortions of the Riemannian energy surfaces induced by time-modulation of the $J_{kl}(t)$ couplings.} We note that the permutation $p_5$ in \eqref{p} can be alternatively realized by modulating the qubit couplings $(J_{12},J_{23})$ instead.

The generators $p_i$ in \eqref{p} form a small non-Abelian permutation group $\cal G$ with 576 elements, 
whose group extension can be read as  
\begin{equation}\label{G}
    {\cal G}=\Big[\left({\cal A}_4\times {\cal A}_4\right):{\cal C}_2\Big]:{\cal C}_2.
\end{equation}
The core of the group $\cal G$ is formed by the direct product of two alternating groups ${\cal A}_4$ of order 12, followed by nested semidirect products of the two cyclic groups ${\cal C}_2$ each of order 2~\cite{MilneBook}.

The action of the group $\cal G$ on the eight elements, represented by the eight NHH eigenvectors, can be described as follows.
  The first direct product ${\cal A}_4\times {\cal A}_4$ represents the independent actions of two alternating groups ${\cal A}_4$, each acting on a disjoint set of four elements out of eight. 
 Next, the cyclic group ${\cal C}_2$ acts on ${\cal A}_4\times {\cal A}_4$  by swapping the two ${\cal A}_4$ subgroups, effectively interchanging the two sets of 4 elements.
 Finally, an outer cyclic group  ${\cal C}_2$ acts on the entire structure, providing an additional global transformation, such as inverting or permuting elements across the entire set.

Geometrically, the elements of the group $\cal G$ can describe the symmetries of chiral three-dimensional polytopes~\cite{Conder2017}. These geometric structures can naturally arise in the form of vertex-figures of higher-dimensional polytopes~\cite{Conder2017,CoxeterBook}.

Clearly, by using the generators $p_i$ in \eqref{p}, one can produce various permutations (576 in total) between the eigenvectors $\psi_i$. For instance, one can readily generate the following group elements of $\cal G$ written as 
\begin{eqnarray}
    p_{6}p_5p_1p_3p_1&=&(\psi_1,\psi_2)(\psi_3,\psi_4)(\psi_5,\psi_6)(\psi_7,\psi_8), \label{p1} \\
    p_1p_5p_1 &=& (\psi_1,\psi_8)(\psi_2,\psi_7)(\psi_3,\psi_6)(\psi_4,\psi_5). \label{p2}
\end{eqnarray}
The group element in \eqref{p1} represents a cyclic shuffling of eigenstates belonging to the two nearest energy manifolds, while the element in \eqref{p2} corresponds to pairwise permutations between eigenstates whose energy levels are `mirror reflected' to each other.

As such, a non-Abelian permutation group $\cal G$, acting on the eigenspace of the NHH, can be constructed through appropriate time modulations of the system parameters in \eqref{param}. This construction utilizes the nontrivial energy topology associated with the EPs of the pseudo-Hermitian Hamiltonian $H$, which effectively suppresses the occurrence of NATs in the system's dynamics. 
{Notably}, these state permutations are topological in nature, {since} the described state-switching protocol does not depend on the shape of the dynamical loops which enclose the EPs.

\section{Discussion}\label{Discussion}
It is evident that by considering larger systems with $N>3$ qubits, more complex permutation groups can be implemented in the system dynamics. {We note that even in the three-qubit system studied, other state permutations may be found by choosing different time-modulating functions of system parameters in~\eqref{param}.}

Importantly, the ability to realize permutation groups in the time-dependent NHH in \eqref{H} enables dynamical state transfer or swapping for any arbitrary state, {since} any state can be expressed as a superposition of the NHH eigenstates. This is in striking contrast with previous studies on dynamical winding around or in the vicinity of EPs, where non-adiabatic transitions prevent bijective mappings over NHH eigenstates due to chiral state switching~\cite{Doppler_2016,Ozdemir2019}.

We would like to emphasize that when the initial state in the protocol described above is a superposition of the NHH eigenstates, the permuted eigenstates in that superposition accumulate an additional phase during state evolution. This phase generally comprises two components: the topological phase, arising from the encirclement of the EPs, and the dynamic phase. 
{Such nontrivial phase accumulation \cite{Garrison_1988, Chu_1989}} can be further utilized in designing phase-dependent quantum protocols, akin to those employed in holonomic computation schemes~\cite{Zhang_2023,Zhang2018_hol,Chen2021_hol}. 

{We also reiterate that the divergent behavior of the spectrum near the EP, as indicated by \eqref{lim}, and associated with it infinite fields $\vec{\bold B}(x,y)$, does not affect the state-switching protocol proposed here. This is because the spectrum remains finite along the EP encircling trajectories. Moreover, as noted earlier {in Section~\ref{sec:eigen}.A}, such divergences can be effectively circumvented by appropriately mapping the NHH onto a higher-dimensional NHH (see also Appendix~\ref{AB}), thus allowing to access such EPs experimentally.}

The NHH of the form in \eqref{H} with effective {complex-valued} 
magnetic fields can be experimentally realized, e.g., through a Naimark dilation method, where the given NHH constitutes a part of a larger Hermitian system~\cite{Gunter2008,Wu2019,Zhang2019_dilation,Huang2022,Liu_enc2021}. Moreover, because the pseudo-Hermiticity of the NHH $\hat H$ [(\eqref{H})]
does not {undergo} spontaneous symmetry breaking, the exact form of the dilated Hermitian Hamiltonian can be readily found~\cite{Huang2022} (see 
Appendix~\ref{Dilation} for details). 
This method can be experimentally implemented using platforms based on nitrogen-vacancy centers in diamond~\cite{Zhang2019_dilation,Liu_enc2021} or superconducting quantum processors~\cite{Dogra2021,Quijandria2018}.

Alternatively, the NHH can be simulated in open quantum optical systems operating in the hard-core boson regime, where a bosonic system can be effectively mapped to a spin system~\cite{Matsubara1956}. For example, in a lossy nonlinear cavity with strongly interacting photons, the system dynamics can be effectively mapped onto those of interacting qubits~\cite{Imamoglu1997}.
In {that} case, the complex magnetic field can act as a complex driving coherent field, which can be emulated using postselection techniques based on $\beta$-dyne detection~\cite{Minganti2022} (see also Appendix~\ref{betadyne} for details).  

Furthermore, the experimental simulation of the NHH can also be performed in classical and quantum photonic setups consisting of {real-space} networks of dissipatively coupled  waveguides, {or even synthetic spaces}, whose coupling topology is identified with the Fock space of the NHH~\cite{Mukherjee2017,Ding2019,Dong2025,Leefmans2022,Parto2023}. 

The described protocol is robust against 
perturbations in the system {that are sufficiently small to ensure that (i) the energy gap opening at the branch cut remains small enough that the state transfer} is still feasible via Landau–Zener–Stückelberg–Majorana transitions~\cite{SHEVCHENKO2010,Ivakhnenko2023}, and (ii) the induced disorder does not trigger {non-adiabatic transitions} in the system dynamics.
Indeed, in general, arbitrary disorder breaks the pseudo-Hermiticity of the NHH in \eqref{H}, leading to a complex-valued spectrum and {to system instabilities associated with this}. 
We elaborate {further} on this in Appendix~\ref{Perturbation}. 

The protocol also remains stable when the modulated interaction terms $J_{kl} \approx 0$ (but not exactly zero) at $t = T/2$ in \eqref{param}, meaning that the different Riemann sheet pairs do not precisely intersect at that moment. This robustness is again attributed to the Landau–Zener–Stückelberg–Majorana transitions between the corresponding energy manifolds, ensuring that the observed state permutations are still achievable~\cite{arkhipov2024a}.
Furthermore, variations in the orbiting trajectory of the magnetic field in the $(x,y)$ plane do not affect the resulting state permutations, as long as the dynamical loop encloses the EPs and does not pass in its close proximity.
{However, we observe that for certain encircling loops and specific states, numerical simulations may exhibit stiffness.  We discuss this difficulty and the way to overcome it in more detail in Appendix~\ref{AG}.}

We note that although in this work we specifically focused on the NHH in the form given in \eqref{H}, other types of NHHs exhibiting various dynamic permutation groups in their eigenspace can also be identified in a similar manner. Our primary aim here was to highlight and describe the very possibility of the existence of such Hamiltonians that enable controlled and stable symmetric state switching in multiqubit systems via dynamical encirclement of EPs.
{However,} the formalism presented here can also be further extended to multiqubit systems with {any} number of qubits. This could open up new possibilities for implementing genuine adiabatic quantum computing in non-Hermitian setups, e.g., realizing {two-qubit quantum protocols in addition to the one- and three-qubit gates presented here}.

\section{Conclusion}
In this work, we have demonstrated that genuine topological quantum state permutations can {indeed} be realized in multiqubit non-Hermitian systems by dynamically encircling EPs, {by converting the Hamiltonian to a pseudo-Hermitian Hamiltonian.} This is accomplished by {adding complex fields to the system and} specifically selecting a subspace 
where the NHH exhibits a real-valued spectrum and hosts a novel type of EPs {defined on hyperboloid manifolds in the system parameter space. We term these exception points 'twisted EPs'.} Dynamically winding around these EPs {allows the} 
suppression of non-adiabatic transitions in the system dynamics, thereby enabling precise and controlled {quantum} state permutations.

As an illustrative example, {in this work} we studied an NHH describing an interacting three-qubit system. We showed that appropriately time-modulating the system parameter space allows {the implementation of} a nontrivial non-Abelian permutation group acting on the eigenstates of the given NHH. Furthermore, the group generators enable symmetric switching between any two eigenstates of the NHH, regardless of the winding direction, {removing the mode chirality imposed by non-adiabatic transitions in conventional non-Hermitian systems.}

The NHHs {with twisted EPs that are proposed and analyzed in this work} can be experimentally realized across various platforms, including solid-state systems, such as superconducting qubits or nitrogen-vacancy centers in diamond using the dilation method~\cite{Dogra2021,Quijandria2018,Zhang2019_dilation,Liu_enc2021}. 
They can also be simulated in the monitored open quantum interacting bosonic systems by performing a photon postselection procedure, or emulated in classical photonic setups through networks of coupled cavities or waveguides~\cite{Minganti2022,Mukherjee2017,Ding2019,Dong2025,Leefmans2022,Parto2023}.

The ability to perform controlled state permutations, while dynamically encircling EPs, introduces new opportunities for quantum state generation and transfer. 
{In particular, the method described here may lead to novel experimental approaches to performing holonomic quantum operations that are based on non-Hermitian quantum physics. Indeed, current non-Hermitian holonomic protocols are severely limited by 
{non-adiabatic transitions between energy eigenstates} and {the losses associated with these}~\cite{Shan2024}. Successful realizations of such approaches are likely to require additional corrective controls to manage spurious transitions when performing operations on quasi-diabatic timescales.}
{In constrast, the present approach exhibits no non-adiabatic losses and can be implemented by standard control methods.}
{The current findings thus} highlight the potential of non-Hermitian systems to advance quantum information and communication protocols, opening up novel avenues for state control and manipulation.

\acknowledgements
I.A. acknowledges support from Air Force Office of Scientific Research (AFOSR) Award No. FA8655-24-1-7376, 
from the Ministry of Education, Youth and Sports of the Czech Republic Grant OP JAC No. CZ.02.01.01/00/22\_008/0004596, and from Grant Agency of the Czech Republic (Project No. 25-15775S).
F.N. is supported in part by:
the Japan Science and Technology Agency (JST)
[via the CREST Quantum Frontiers program Grant No. JPMJCR24I2,
the Quantum Leap Flagship Program (Q-LEAP), and the Moonshot R\&D Grant Number JPMJMS2061],
and the Office of Naval Research (ONR) Global (via Grant No. N62909-23-1-2074).
S.K.O., P.L. and K.B.W. 
acknowledge support of 
AFOSR Multidisciplinary University Research Initiative (MURI)
Award on Programmable systems with non-Hermitian quantum dynamics
(Award No. FA9550-21-1-0202).

\appendix

\section{Pseudo-Hermitian symmetry of the non-Hermitian Hamiltonian in Eq.~(1)}\label{Metric}
Here we provide further details on {the} pseudo-Hermiticity of the NHH in \eqref{H} in the main text. 
{Specifically,} we want to show that one can find 
a Hermitian operator $\eta$ for which one always has
\begin{eqnarray}\label{A1}
    \eta\hat H\eta^{-1}=\hat H^{\dagger},
\end{eqnarray}
The operator $\eta$ can be chosen as 
\begin{eqnarray}\label{eta}
    \eta=\bigotimes\limits_k^N\xi, \quad \xi=\sum\limits_{i=1}^2|\nu_i\rangle\langle\nu_i|.
\end{eqnarray}
{where the states $|\nu_i\rangle$ are the left eigenvectors of the single qubit Hamiltonian $H_q=B_x\sigma_x+B_y\sigma_y$, with $B_{x,y}$ given by \eqref{H} - \eqref{alpha}, i.e., $H_q^{\dagger}|\nu_i\rangle\equiv |\nu_i\rangle$.}
That is, the operator $\xi$ is a metric operator of the qubit Hamiltonian $H_q$. We recall that the left eigenvectors $\nu_{1,2}$ are complex conjugates of the right eigenvectors in \eqref{psi}. 

Because of this {property}, the non-interacting part ($\textbf{J}=0$) of the NHH in \eqref{H}, namely, $H_{\rm non}=\sum\limits_k^N\vec{{\boldsymbol\sigma}}^k\vec{\bold B}_{||}^k$, automatically fulfills the condition in \eqref{A1}, since the operator $\eta$ is a metric operator of this non-interacting term. 

On the other hand, the interacting term $H_{\rm int}=\sum J_{kl}\hat\sigma_z^k\hat\sigma_z^l$ is invariant under the action of the operator $\eta$ in \eqref{eta}, i.e., $\eta\hat H_{\rm int}\eta^{-1}=H_{\rm int}$, which can be checked by straightforward calculations. 

{Consequently, we obtain}
\begin{eqnarray}
    \eta\hat H\eta^{-1}=\eta\hat H_{\rm non}\eta^{-1}+\eta\hat H_{\rm int}\eta^{-1}=\hat H_{\rm non}^{\dagger}+\hat H_{\rm int}. \nonumber \\
\end{eqnarray}
Because 
$\hat H_{\rm int}$ is Hermitian by definition, one {then} arrives at the pseudo-Hermitian condition in \eqref{A1}. 

\section{Resolving the problem of exceptional points at infinity for single qubits in \eqref{lim}}\label{AB}
Here, we briefly address the apparent issue of spectral divergence at EPs in the non-interacting qubits NHH, according to \eqref{lim}. This problem can be resolved using the framework of complex projective spaces~\cite{NakaharaBook}, which provides an elegant method to handle infinities. In particular, what appears as an infinity in a given space may correspond to a well-defined point in a higher-dimensional space.

To put it more formally, the procedure begins with identifying a higher-dimensional NHH that can be associated with a Grassmann manifold, which generalizes the concept of a projective space. Then, a particular class of this Grassmannian would corresponds to the lower-dimensional NHH, specifically the single-qubit NHH discussed in Sec.~\ref{nonint}.

Such a higher-dimensional $3\times 3$ NHH can be introduced as follows
\begin{eqnarray}\label{H'}
    H'=\begin{pmatrix}
        \gamma\cot\phi & \gamma & 0 \\
        \gamma & -\gamma\cot\phi & 0 \\
        0 & 0 & \dfrac{\sin\left({\rm Im}[\phi]\right)}{\sin\phi}
    \end{pmatrix},
\end{eqnarray}
where $\gamma =\Delta\sin\left({\rm Re}[\phi]\right)$.
The first two diagonal elements of the NHH are continuous functions on the plane $(x,y)$. 
The third element is also a continuous function except the interval $[x=0,|y|\leq1]$, where it becomes rather a piecewise function, i.e.,  it is not differentiable but remains finite on that interval. 

The NHH possesses an EP at the same point $(x=0,y=1)$ as the NHH of a single qubit considered in Sec.~\ref{nonint}. Moreover, the spectrum of the NHH $H'$ in \eqref{H'} is finite in any given finite region $(x,y)$ including the EP. These eigenvalues read
\begin{eqnarray}\label{lambda1}
    \lambda_{1,2}=\pm \gamma(\sin\phi)^{-1}, \quad \lambda_3=\sin\left({\rm Im}[\phi]\right)(\sin\phi)^{-1}.
\end{eqnarray}
The second-order EP is characterized by the coalescence of two eigenvectors associated with the degenerate eigenvalues 
\begin{eqnarray}\label{lambda}
    \lambda_1^{\rm EP}=\lambda_2^{\rm EP}=0.
\end{eqnarray}
Notably, while the third eigenvalue is also $\lambda_3=0$ at the EP, its corresponding eigenvector remains independent of the other two eigenvectors.

\begin{figure*}[t!]
    \includegraphics[width=0.98\textwidth]{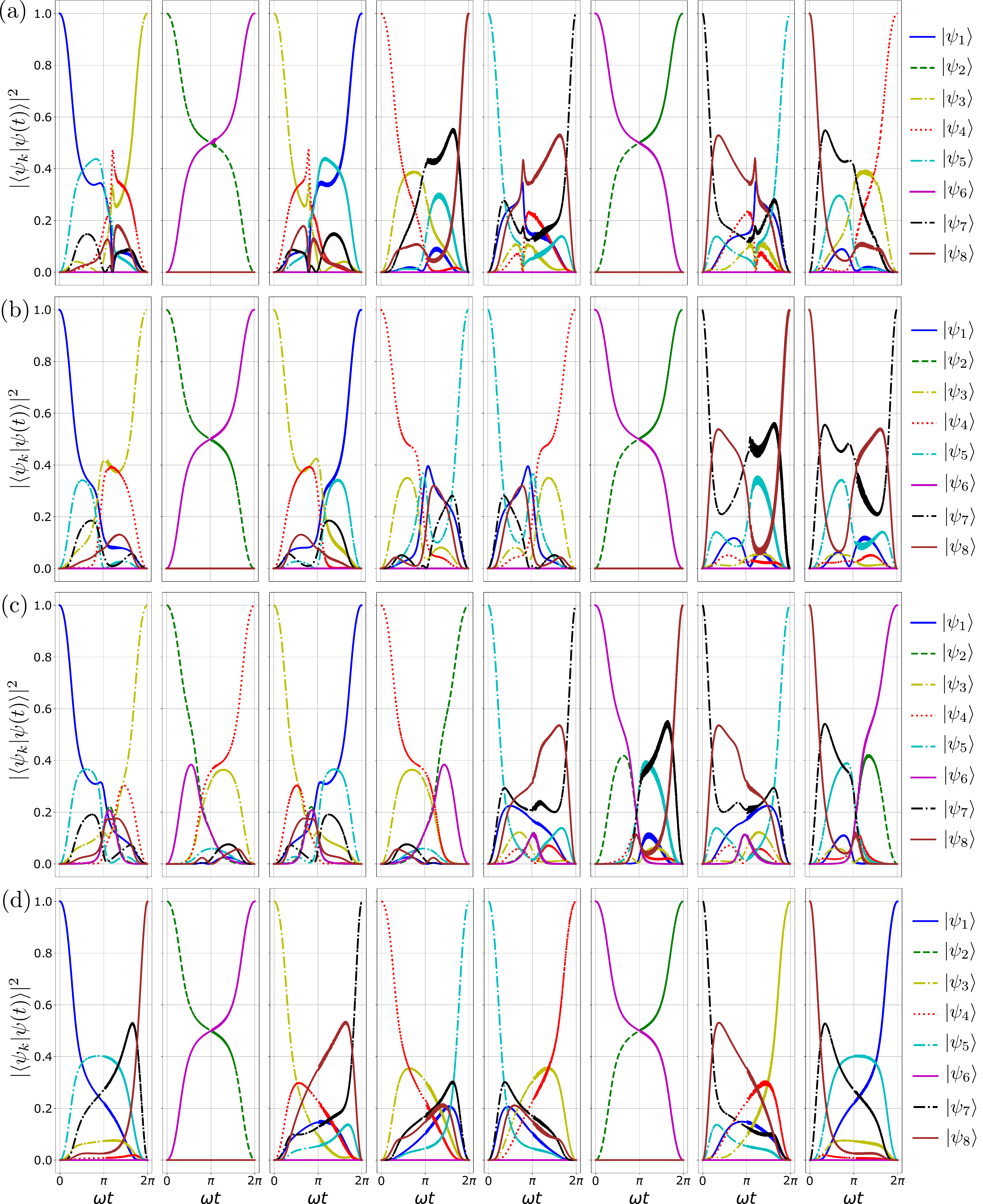}
    \caption{This Figure is an extension of {Figs.~\ref{fig4}(d) and \ref{fig5}(d),} i.e., the calculated state transfer fidelity $|\langle\psi_k|\psi(t)\rangle|^2$ between evolving state $|\psi(t)\rangle$ and one of the eight initial eigenstates $|\psi_k\rangle$ of the three-qubit NHH. The panels in the plot are related to different combinations of the modulated system parameters $J_{kl}$ in \eqref{param}, and which thus correspond to permutation generators in \eqref{p}. The panels (a),(b),(c) and (d) correspond to the permutation generators $p_3$, $p_4$, $p_5$ and $p_6$, respectively. The semi-minor axis and semi-major axis of the encircling ellipse are $r_x=1$ and $r_y=2$, respectively. The remaining system parameters are the same as in {Figs.~\ref{fig4}(d) and \ref{fig5}(d).}
    }
    \label{S1}
\end{figure*}

\begin{figure*}[t!]
    \includegraphics[width=0.99\textwidth]{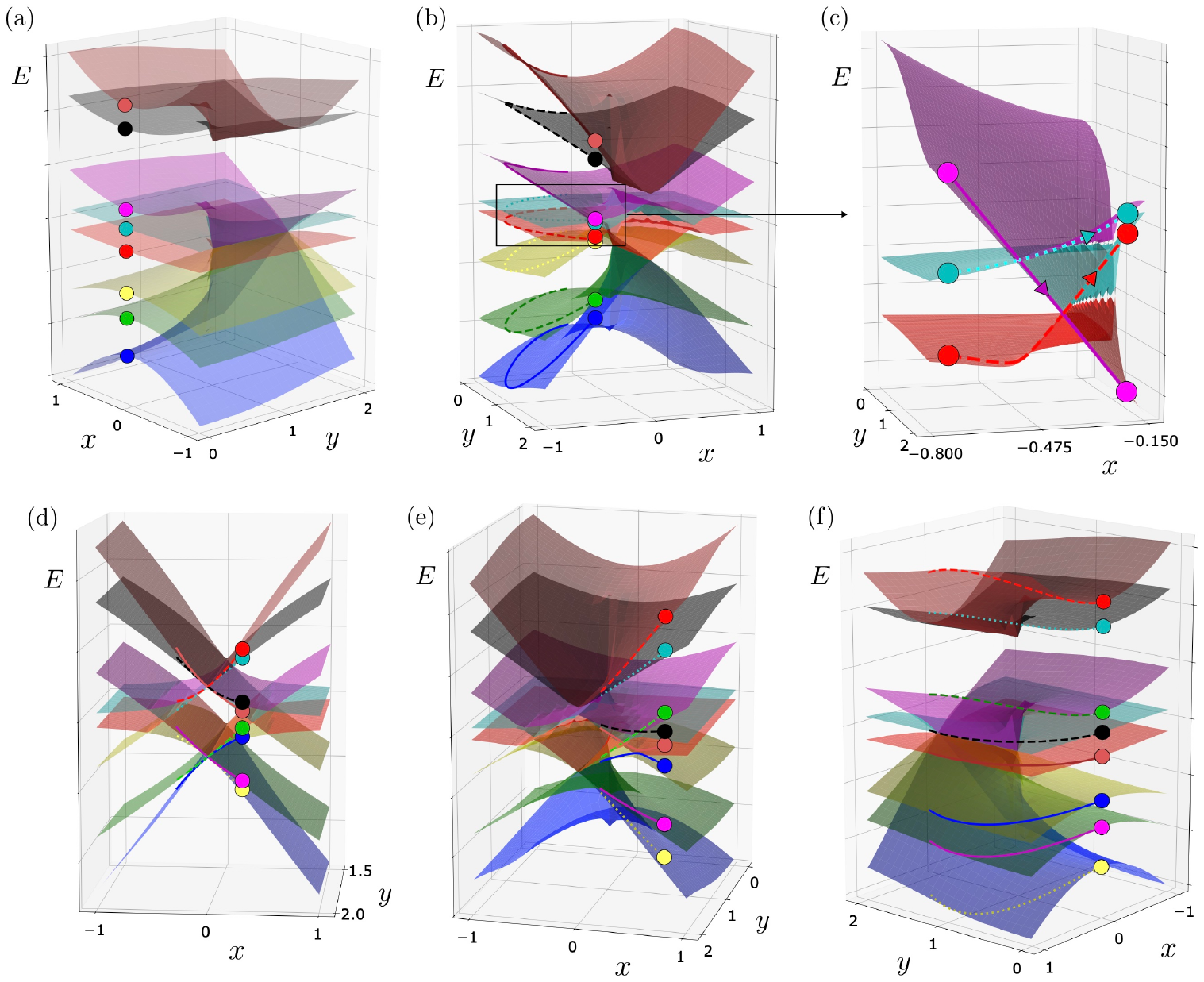}
    \caption{
     {Schematic representation of implementing an element $p_3$ of the permutation group $\cal G$ in \eqref{p} in the eigenspace of a three-qubit system governed by the NHH in \eqref{H}, while dynamically  spiraling EPs over a winding period $T$. The system parameters are modulated according to \eqref{param}. This figure also corresponds to the fidelity graph shown in \figref{S1}(a). (a)  Initial eigenstates at $t=0$, depicted as colored spheres, with each color corresponding to a certain Riemann energy manifold [same as in Figs.~\ref{fig4}(a) and \ref{fig5}(a)]. (b) The position of evolving eigenstates at time $0<t<T/2$. (c) Zoomed in incaption from panel (b), showing the evolution of the eigenstates $\psi_4$ (red sphere), $\psi_5$ (cyan sphere) and $\psi_6$ (magenta sphere), which pass through various Riemann energy manifolds. (d,e,f) The evolution of the eigenstates at times $t=T/2$, $T/2<t<T$, and $t=T$, respectively. See the main text in Appendix~\ref{Permutations1} for more details.} 
    }\label{S2}
\end{figure*}

\begin{figure*}[t!]
    \includegraphics[width=0.99\textwidth]{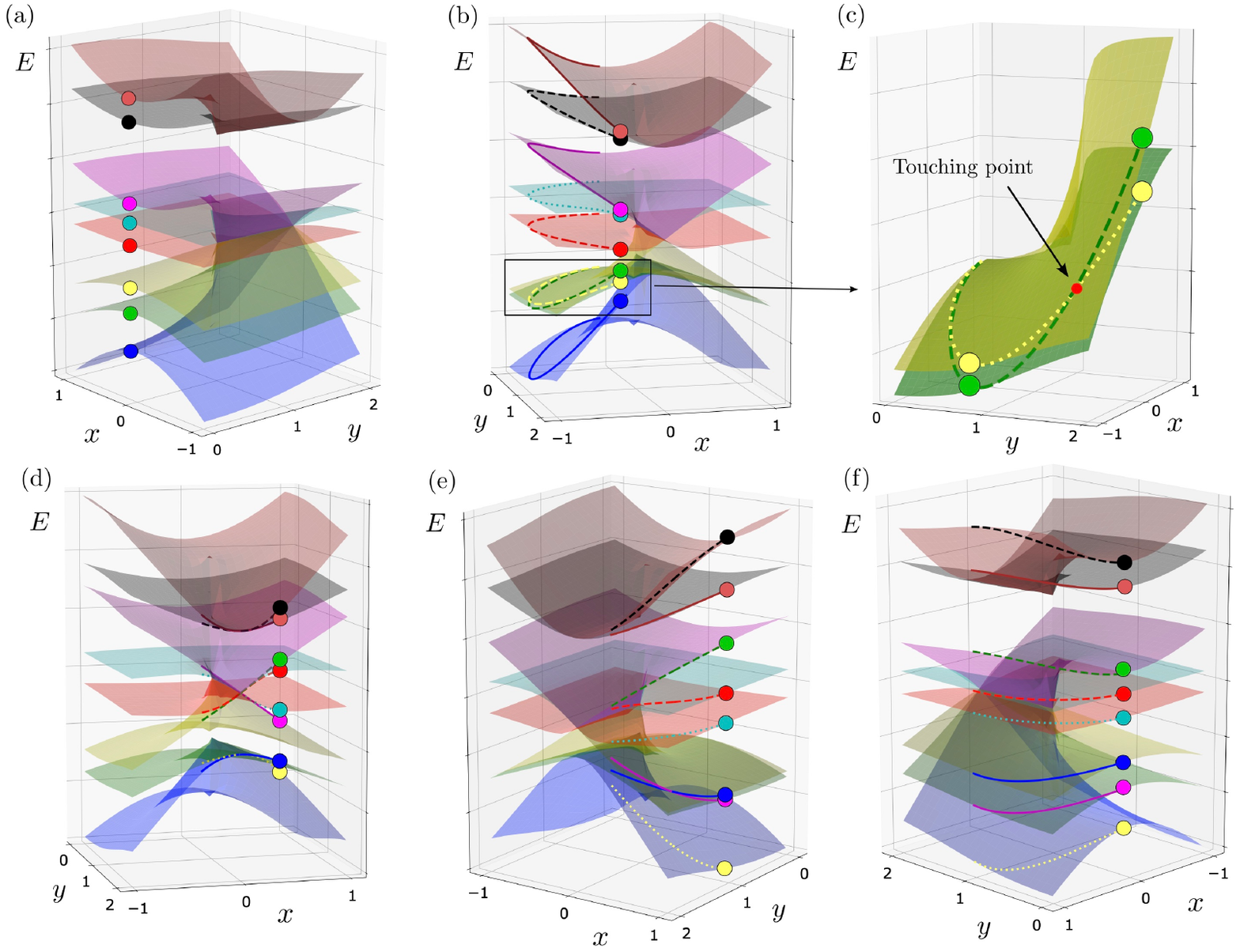}
    \caption{
     {Schematic representation of implementing an element $p_4$ of the permutation group $\cal G$ in \eqref{p} in the eigenspace of a three-qubit system governed by the NHH in \eqref{H}, while dynamically  spiraling EPs over a winding period $T$. The system parameters are modulated according to \eqref{param}. This figure also corresponds to the fidelity graph shown in \figref{S1}(b). (a)  Initial eigenstates at $t=0$, depicted as colored spheres, with each color corresponding to a certain Riemann energy manifold [same as in Figs.~\ref{fig4}(a) and \ref{fig5}(a)]. (b) The position of evolving eigenstates at time $0<t<T/2$. (c) Zoomed in incaption from panel (b), showing the evolution of the eigenstates $\psi_2$ (green sphere) and $\psi_3$ (yellow sphere), whose trajectories along the corresponding Riemann manifolds are exchanged at the touching point of these energy sheets. (d,e,f) The evolution of the eigenstates at times $t=T/2$, $T/2<t<T$, and $t=T$, respectively. See the main text in Appendix~\ref{Permutations2} for more details.}
    }\label{S3}
\end{figure*}
\begin{figure*}[t!]
    \includegraphics[width=0.85\textwidth]{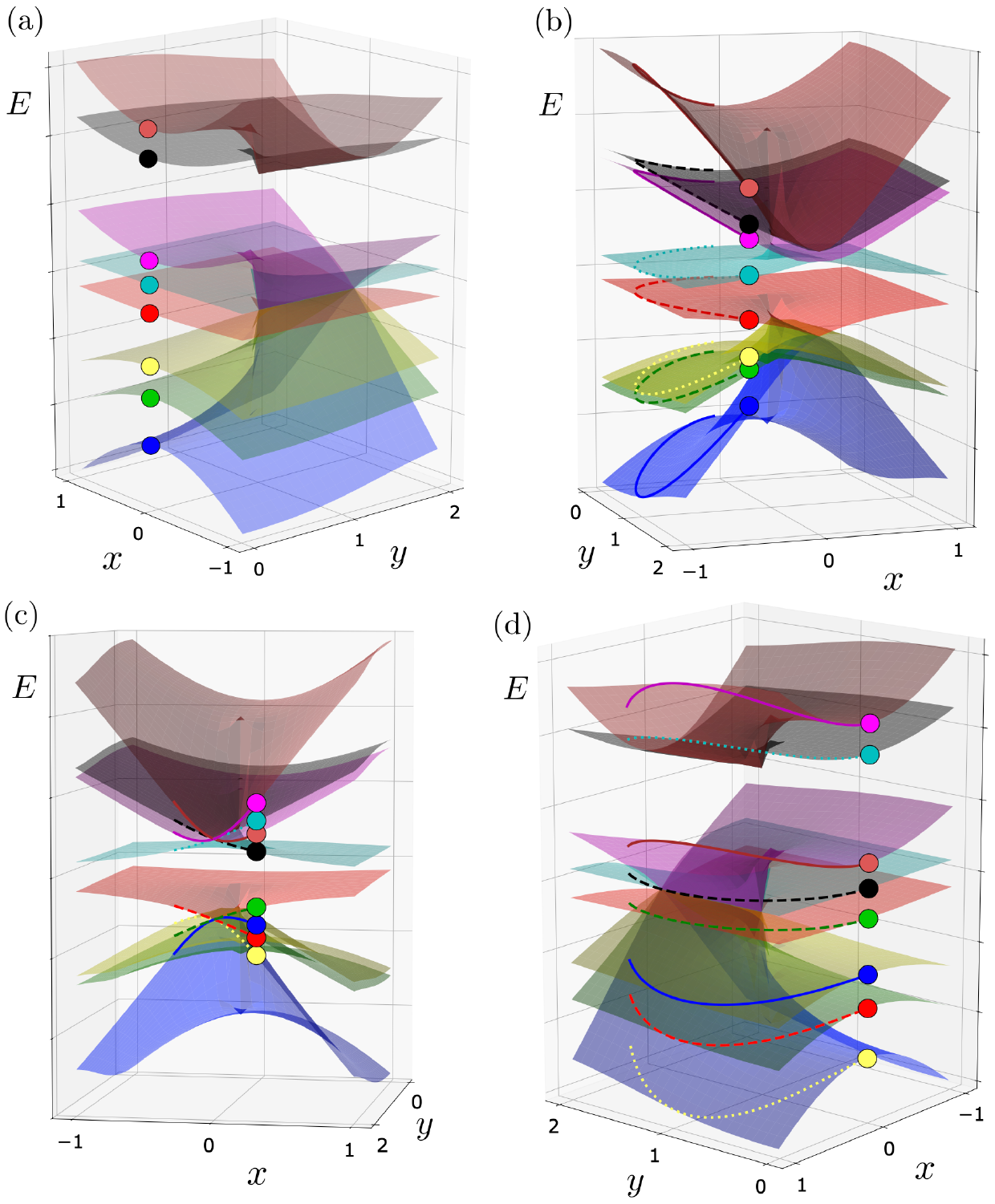}
    \caption{
    {Schematic representation of implementing an element $p_5$ of the permutation group $\cal G$ in \eqref{p} in the eigenspace of a three-qubit system governed by the NHH in \eqref{H}, while dynamically  spiraling EPs over a winding period $T$. The system parameters are modulated according to \eqref{param}. This figure also corresponds to the fidelity graph shown in \figref{S1}(c). (a)  Initial eigenstates at $t=0$, depicted as colored spheres, with each color corresponding to a certain Riemann energy manifold [same as in Figs.~\ref{fig4}(a) and \ref{fig5}(a)]. (b,c,d) The position of the evolving eigenstates at times $0<t<T/2$, $t=T/2$, and $t=T$, respectively. See the main text in Appendix~\ref{Permutations3} for more details.}
    }\label{S4}
\end{figure*}

\begin{figure*}[t!]
    \includegraphics[width=0.85\textwidth]{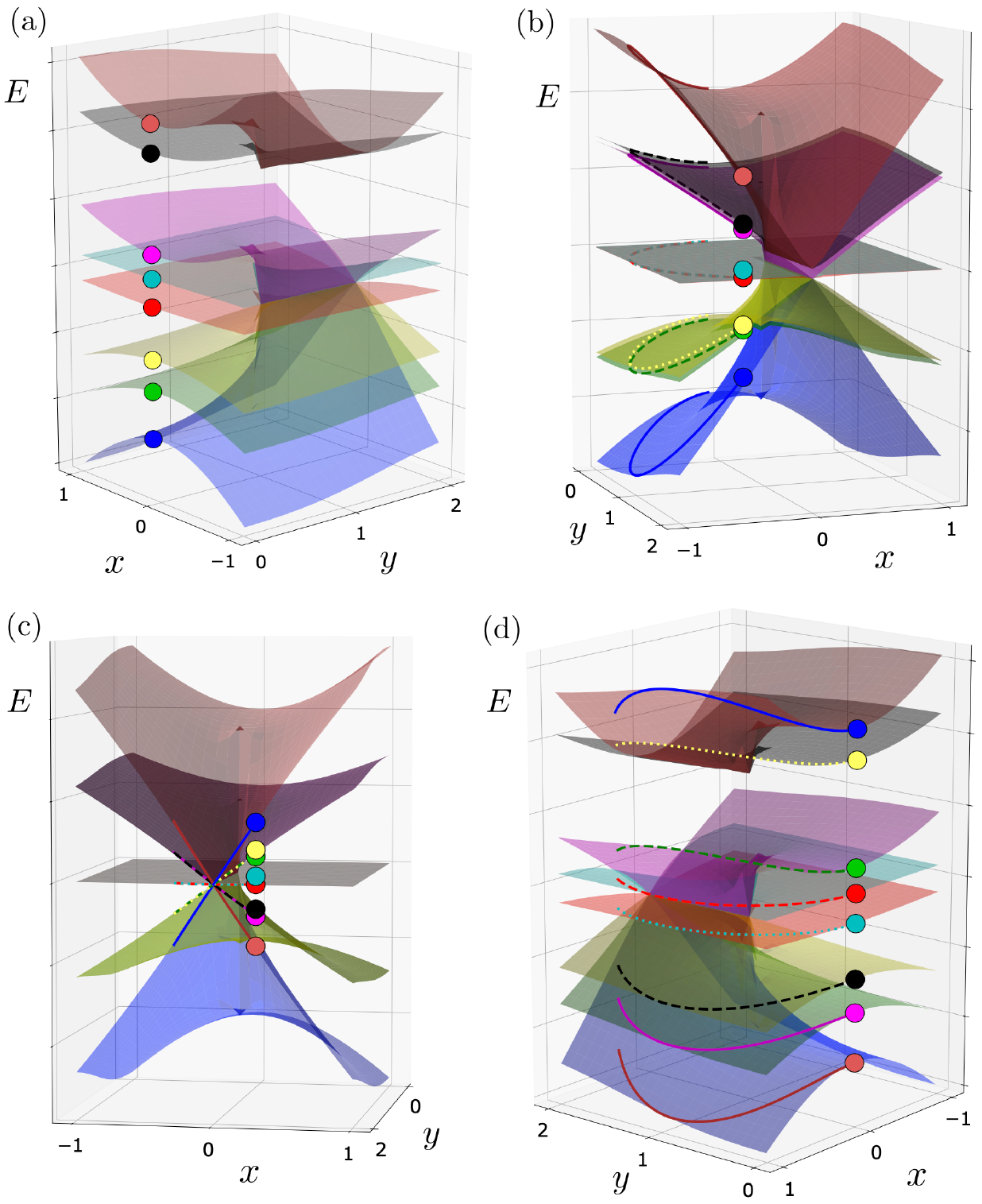}
    \caption{
    {Schematic representation of implementing an element $p_6$ of the permutation group $\cal G$ in \eqref{p} in the eigenspace of a three-qubit system governed by the NHH in \eqref{H}, while dynamically  spiraling EPs over a winding period $T$. The system parameters are modulated according to \eqref{param}. This figure also corresponds to the fidelity graph shown in \figref{S1}(d). (a)  Initial eigenstates at $t=0$, depicted as colored spheres, with each color corresponding to a certain Riemann energy manifold (same as in Figs.~\ref{fig4} and \ref{fig5}). (b,c,d) The position of the evolving eigenstates at times $0<t<T/2$, $t=T/2$, and $t=T$, respectively. See the main text in Appendix~\ref{Permutations4} for more details.}
    }\label{S5}
\end{figure*}
Now, we can associate subspaces of this $3\times 3$ NHH with a Grassmanian ${G}_{1,3}$, which is the projective space ${\mathbb CP}^2$~\cite{NakaharaBook}. In the case under consideration, we define the projective space ${\mathbb CP}^2$ with the coset $cH'$, where $c\in {\mathbb C}-\{0\}$. 

One of the charts of the this projective space can be defined as $\dfrac{\sin\phi}{\sin\left({\rm Im}[\phi]\right)} H'$, obtained by normalizing $H'$ by its last diagonal element. The representative of this chart with inhomogeneous coordinates can then be expressed via the $2 \times 2$ matrix 
$H = \alpha \begin{pmatrix}
    \cos\phi & \sin\phi \\
    \sin\phi & -\cos\phi
\end{pmatrix}$,
which corresponds to the NHH of the single qubit discussed in Sec.~\ref{nonint}.

In other words, through this projection, the well-defined EP in \eqref{lambda} is sent to infinity in \eqref{lim}. The latter corresponds to the situation when the projection of the EP in ${\mathbb C}^3$, determined by the $3\times 3$ matrix $H'$ to the plane, defined by the $2\times 2$ matrix $H$, is parallel to this plane. 

Thus, the projective space associated with the NHH $H$ provides a framework to effectively map the EP in \eqref{lambda} to infinity, according to \eqref{lim}.
We reserve the discussion of experimental implementation of this mathematical approach for future studies.

\section{Calculated fidelity for other state permutations according to \eqref{p}}\label{Permutations}
In {Figs.~\ref{fig4}(d) and \ref{fig5}(d)} we plotted the fidelity $|\langle\psi_k|\psi(t)\rangle|^2$ between the evolving state $|\psi(t)\rangle$ and one of the eight initial eigenstates $|\psi_k\rangle$ of the three-qubit NHH that correspond to the first generators of the permutation group $\cal G$ in \eqref{p}. For completeness,  here in \figref{S1} we plot the fidelity of the remaining permutation generators, thus confirming their form in \eqref{p}. {Visualizations of he state dynamics corresponding to these permutations are shown in 
Figs. \ref{S2} - \ref{S5} and
the various eigenstate flips produced by the permutation generators $p_i$ are summarized} in  Table~\ref{table}. From this Table it is clear that one can always find such a combination of modulated system parameters in \eqref{param} that enables a state transfer between any two given eigenstates in one or more full cycles $T$.

Below we analyze in detail the evolution of eigenstates corresponding to each panel in \figref{S1}. 
\subsection{Visualization of the eigenstates evolution corresponding to the element $p_3$ of the permutation group $\cal G$ in \eqref{p}}\label{Permutations1}
We begin with the state dynamics illustrated in \figref{S2}, whose fidelity evolution is shown in \figref{S1}(a), corresponding to the permutation group element $p_3$ of $\cal G$ defined in \eqref{p}.

At $t=0$, the initial eigenstates $\psi_i$ occupy their respective Riemann eigenenergy sheets $E_i$ in the Hermitian phase of the non-Hermitian Hamiltonian $\hat{H}$, where the complex magnetic field $\boldsymbol{\vec{B}} = 0$, i.e., $(x=0, y=0)$ in \figref{S2}(a). The eigenstates $\psi_i$ are represented as colored spheres, consistent with Figs.~\ref{fig4} and \ref{fig5}: $\psi_1$ (blue), $\psi_2$ (green), $\psi_3$ (yellow), $\psi_4$ (red), $\psi_5$ (cyan), $\psi_6$ (magenta), $\psi_7$ (black), and $\psi_8$ (brown).

As the system parameters are modulated according to~\eqref{param}, including the time-dependent couplings $J_{13}(t)$ and $J_{23}(t)$, the Riemann sheets begin to deform, {showing increasing bending,} and {also shift} along the $z$-axis, as shown in \figref{S2}(b). At an intermediate time $0 < t < T/2$ within the range $[-1 < x < 0,\ 0 < y < 2]$, the energy manifolds $E_4$, $E_5$, and $E_6$ undergo mutual permutation, according to the adopted convention in \eqref{order}, [see \figref{S2}(c), which is a zoom-in incaption of panel (b)]. Consequently, the associated eigenstates are reshuffled: $\psi_4$ begins evolving along $E_5$, $\psi_5$ follows $E_6$, and $\psi_6$ follows $E_4$.

At the midpoint of the evolution, $t = T/2$ ($x = 0$, $y = 2$), several Riemann sheets become coupled: 
$E_1 \leftrightarrow E_3$, $E_2 \leftrightarrow E_4$, $E_5 \leftrightarrow E_8$, and $E_6 \leftrightarrow E_7$ [\figref{S2}(d)]. These couplings cause the eigenstates to switch to different sheets: $\psi_1 \rightarrow E_3$, $\psi_2 \rightarrow E_4$, $\psi_3 \rightarrow E_1$, $\psi_4$ (on $E_5$) $\rightarrow E_8$, $\psi_5$ (on $E_6$) $\rightarrow E_7$, $\psi_6$ (on $E_4$) $\rightarrow E_2$, $\psi_7 \rightarrow E_6$, and $\psi_8 \rightarrow E_5$.

Subsequently, during the second half of the cycle, for $T/2 < t < T$, shown in \figref{S2}(e) an identical permutation among $E_4$, $E_5$, and $E_6$ occurs, mirroring the earlier rearrangement in \figref{S2}(c). This time, the permutation involves the states $\psi_2$, $\psi_7$, and $\psi_8$.

As a result of the complete encirclement around the EPs, as depicted in \figref{S2}(f), one realizes the permutation element:
\begin{eqnarray}
    p_3=(\psi_1,\psi_5)(\psi_2,\psi_6)(\psi_3,\psi_7)(\psi_4,\psi_8),
\end{eqnarray}
according to \eqref{p}.

\subsection{Visualization of the eigenstates evolution corresponding to the element $p_4$ of the permutation group $\cal G$ in \eqref{p}}\label{Permutations2}
By modulating the magnetic field $\boldsymbol{\vec{B}}(x,y)$ along with the time-dependent qubit coupling $J_{12}(t)$, according to \eqref{param}, one can realize the eigenstate permutation associated with the group element $p_4$ of the permutation group $\cal G$ defined in~\eqref{p}. The resulting evolution of the eigenstates corresponding to $p_4$ is illustrated in \figref{S3}.

As in \figref{S2}(a), the initial eigenstates at $t = 0$ are aligned with their respective Riemann energy sheets in the Hermitian phase $\boldsymbol{\vec{B}}(x,y)=0$, as shown in \figref{S3}(a). At an intermediate stage, $0 < t < T/2$, in the domain $[-1 < x < 0,\ 0 < y < 2]$, the deformation of the energy sheets $E_2$ and $E_3$ due to the modulation of $J_{12}(t)$ may cause these surfaces to come into proximity, where they touch each other [see \figref{S3}(c), which is a zoom-in of {the rectangular region highlighted} in panel (b)]. As a result, the corresponding eigenstates undergo a permutation: $\psi_2$ begins evolving along $E_3$, while $\psi_3$ transitions to $E_2$.

At the halfway point of the parameter cycle, $t = T/2$ ($x = 0,\ y = 2$), several Riemann sheets become coupled, though with a different pairing compared to the previous case. The coupled energy surfaces are: $E_1 \leftrightarrow E_2$, $E_3 \leftrightarrow E_6$, $E_4 \leftrightarrow E_5$, and $E_7 \leftrightarrow E_8$ [\figref{S3}(d)]. These couplings induce the following transitions among the eigenstates: $\psi_1 \rightarrow E_2$, $\psi_2$ (on $E_3$) $\rightarrow E_6$, $\psi_3$ (on $E_2$) $\rightarrow E_1$, $\psi_4 \rightarrow E_5$, $\psi_5 \rightarrow E_4$, $\psi_6 \rightarrow E_3$, $\psi_7 \rightarrow E_8$, and $\psi_8 \rightarrow E_7$.

In the second half of the encirclement, for $T/2 < t < T$, a similar exchange between $E_2$ and $E_3$ takes place, mirroring the earlier scenario in \figref{S3}(c). This leads to an additional permutation between $\psi_1$ and $\psi_6$.

At the end of the full cycle ($t = T$), the final configuration of the eigenstates corresponds to the permutation element:
\begin{equation}
p_4 = (\psi_1,\psi_3)(\psi_2,\psi_6)(\psi_4,\psi_5)(\psi_7,\psi_8),
\end{equation}
when compared to the initial state ordering.

\subsection{Visualization of the eigenstates evolution corresponding to the element $p_5$ of the permutation group $\cal G$ in \eqref{p}}\label{Permutations3}
The implementation of the eigenstate permutation $p_5$ from the group $\cal G$ requires, in addition to modulating the magnetic field $\boldsymbol{\vec{B}}(x,y)$, the simultaneous modulation of either the qubit couplings $J_{12}(t)$ and $J_{13}(t)$, or the pair $J_{12}(t)$ and $J_{23}(t)$, in accordance with Eqs.~(\ref{param}) and (\ref{p}). We illustrate the corresponding state evolution in \figref{S4}.

In contrast to the permutations $p_3$ and $p_4$, no eigenstate transitions occur during the intermediate intervals $0 < t < T/2$ or $T/2 < t < T$ [see also \figref{S4}(b)]. Instead, the state evolution is entirely governed by the couplings at the midpoint of the winding cycle, $t = T/2$. At this point, the following Riemann sheet couplings are observed: $E_1 \leftrightarrow E_3$, $E_2 \leftrightarrow E_4$, $E_5 \leftrightarrow E_7$, and $E_6 \leftrightarrow E_8$ [see \figref{S4}(c)]. These couplings induce the corresponding eigenstate transitions: $\psi_1 \rightarrow E_3$, $\psi_2\rightarrow E_4$, $\psi_3\rightarrow E_1$, $\psi_4 \rightarrow E_2$, $\psi_5 \rightarrow E_7$, $\psi_6 \rightarrow E_8$, $\psi_7 \rightarrow E_5$, and $\psi_8 \rightarrow E_6$.
This new configuration of eigenstates remains unchanged for the rest of the evolution until the end of the winding period $t = T$, thus leading to the permutation element $p_5$ defined in~\eqref{p}:
\begin{equation}
p_5 = (\psi_1,\psi_3)(\psi_2,\psi_4)(\psi_5,\psi_7)(\psi_6,\psi_8).
\end{equation}

\subsection{Visualization of the eigenstates evolution corresponding to the element $p_6$ of the permutation group $\cal G$ in \eqref{p}}\label{Permutations4}
To realize the eigenstate permutation $p_6$, one must simultaneously modulate the magnetic field $\boldsymbol{\vec{B}}(x,y)$ and all three qubit couplings $J_{12}(t)$, $J_{13}(t)$, and $J_{23}(t)$, according to Eqs.~(\ref{param}) and (\ref{p}). The corresponding eigenstate dynamics are shown in \figref{S5}.

As in the case of the permutation element $p_5$, the final configuration of eigenstates is solely determined by the couplings between Riemann sheets at the midpoint of the evolution, $t = T/2$, induced by the modulating qubit couplings. Specifically, \figref{S5}(c) shows that the following energy surfaces become coupled and degenerate: $E_1 \leftrightarrow E_8$, $E_2 \leftrightarrow E_6$, $E_3 \leftrightarrow E_7$, and $E_4 \leftrightarrow E_5$. These couplings result in the corresponding transitions between eigenstates: $\psi_1 \rightarrow E_8$, $\psi_2\rightarrow E_6$, $\psi_3\rightarrow E_7$, $\psi_4 \rightarrow E_5$, $\psi_5 \rightarrow E_4$, $\psi_6 \rightarrow E_2$, $\psi_7 \rightarrow E_3$, and $\psi_8 \rightarrow E_1$.
This evolution leads to the realization of the permutation element $p_6$ of the group $\cal G$, given by:
\begin{equation}
p_6 = (\psi_1,\psi_8)(\psi_2,\psi_6)(\psi_3,\psi_7)(\psi_4,\psi_5).
\end{equation}
\color{black}
\begin{table*}[t!]
\caption{Table of the state permutations according to \eqref{p}. Starting from an initial system state $\psi_k$ (listed in the first column), one can transition to any target eigenstate $\psi_j$ (listed in the first row) by selecting an appropriate trajectory (expressed via a permutation generator $p_i$ in \eqref{p})  that winds or spirals around the exceptional point.}
\begin{tabular}{|c||*{8}{>{\centering\arraybackslash}m{5em}|}}
\hline
\backslashbox{$\psi_{\rm in}$}{$\psi_{\rm out}$}
&$\psi_1$&$\psi_2$&$\psi_3$&$\psi_4$&$\psi_5$&$\psi_6$&$\psi_7$&$\psi_8$\\\hline\hline
{$\psi_1$} &  &$p_2$&$p_3,p_4,p_5$& & $p_1$ & & &$p_6$\\\hline
$\psi_2$ &$p_2$& &&$p_5$& &$p_1,p_3,p_4,p_6$ & &\\\hline
$\psi_3$ & $p_3,p_4,p_5$ & & & $p_1$ & & $p_2$ & $p_6$ &\\\hline
$\psi_4$ & &$p_5$ &$p_1$& & $p_2,p_4$ & & & $p_3$\\\hline
$\psi_5$ &$p_1$& & & $p_2,p_4$ &  &  & $p_3,p_5$ &\\\hline
$\psi_6$ & & $p_1,p_3,p_4,p_6$ &$p_2$& & & & & $p_5$\\\hline
$\psi_7$ & & &$p_6$& & $p_3,p_5$ & & &$p_1,p_2,p_4$\\\hline
$\psi_8$ &$p_6$& & &$p_3$& & $p_5$ & $p_1,p_2,p_4$ &\\\hline
\end{tabular}\label{table}
\end{table*}

\section{Realizing the non-Hermitian Hamiltonian in \eqref{H} via the dilation method}\label{Dilation}
In this section, we review the key principles of the dilation method~\cite{Gunter2008}, which can be utilized for the experimental realization of the NHH in \eqref{H}.

The core idea of the dilation method is to embed the Hilbert space of a given NHH into a larger space, where it becomes part of an extended Hermitian Hamiltonian.
Consider the Schrödinger equation for a system governed by a NHH $H$ acting on an $n$-dimensional vector $|\psi\rangle$
\begin{eqnarray}\label{undH}
    i\partial_t|\psi(t)\rangle=H|\psi(t)\rangle.
\end{eqnarray}
To achieve that embedding one can define the following extended $2n$-dimensional state vector 
\begin{eqnarray}\label{extvec}
    |\Psi\rangle=\begin{pmatrix}
    |\psi\rangle \\
    D |\psi\rangle
\end{pmatrix},
\end{eqnarray}
where $D$ is a certain matrix to be defined later.
The dynamics of this dilated vector $|\Psi\rangle$ is determined by a Hermitian Hamiltonian $\mathbf{H}$
\begin{eqnarray}\label{dilH}
    i\partial_t|\Psi(t)\rangle={\mathbf{H}}|\psi(t)\rangle, \quad {\mathbf{H}}={\mathbf{H}}^{\dagger}.
\end{eqnarray}
By presenting this Hermitian Hamiltonian in the block matrix form ${\mathbf{H}}=\begin{pmatrix}
    h_{1} & h_2 \\
    h_2^{\dagger} & {h}_3
\end{pmatrix}$, and combining Eqs.~(\ref{undH})--(\ref{dilH}), one arrives to a system of differential equations
\begin{eqnarray}\label{sys}
    h_1 + h_2D&=&H, \nonumber \\
    h_2^{\dagger}+h_3D&=&i\dot{D}+DH.
\end{eqnarray}
From the system of equations in \eqref{sys} one attains a formal solution for the matrix $h_1$
\begin{eqnarray}
    h_1 = H +i\dot{D}^{\dagger}D-H^{\dagger}D^{\dagger}D+D^{\dagger}h_3D.
\end{eqnarray}
By exploiting the property $h_1=h_1^{\dagger}$, one arrives to the equation
\begin{eqnarray}\label{metric}
    i\partial_t\eta=H^{\dagger}\eta-\eta H,
\end{eqnarray}
where we denoted $\eta = I+D^{\dagger}D$.

The equation \eqref{metric} is precisely the equation for the metric of the NHH $H$~\cite{Ju2022}. Indeed, by construction the $\eta$ is positive definite, and can be explicitly selected as given in \eqref{eta}. Because of the pseudo-Hermiticity of the NHH $H$, the metric $\eta$ in \eqref{metric} is always `constant' when written in the instantaneous left eigenbasis of the NHH.  Moreover, with the choice in \eqref{eta}, the operator $(\eta-I)$ is non-negative, which immediately allows to find the ancillary matrix $D$ as
\begin{eqnarray}\label{D}
    D = U\sqrt{\eta-I},
\end{eqnarray}
where $U$ is an arbitrary unitary matrix.

We stress that even for a time-dependent NHH, the metric retains the form given in \eqref{eta}, owing to the pseudo-Hermitian symmetry of the NHH in the whole system parameter space $(x,y,J)$.
As such, one can readily construct a dilated Hermitian Hamiltonian whose subspace is constituted by the NHH from the knowledge of the metric $\eta$ in \eqref{eta}.

\section{Realizing the non-Hermitian Hamiltonian in \eqref{H} via the $\beta$-dyne detection method}\label{betadyne}
{The NHH of} \eqref{H} can also be experimentally implemented in the $\beta$-dyne detection scheme~\cite{Minganti2022}. This method allows introducing complex driving fields in the system's Hamiltonian. That approach is based on a specific homodyne-detection monitoring scheme of a quantum system surrounded by a Markovian environment into which the system's excitations are leaking out. 

To understand the main idea of the approach, let us first consider a master equation, in the Lindblad form, which defines the dynamics of an open quantum system~\cite{BreuerBookOpen} 
\begin{equation}\label{rho}
    \frac{\partial \rhot}{\partial t} = -i \left[\hat{H}_{\rm eff}, \rhot \right] + \sum_{\mu} \gamma_\mu\hat{a}_\mu \rhot \hat{a}_\mu^{\dagger}.
\end{equation}
In \eqref{rho}, the Hermitian operator $\hat\rho$ is the system's density matrix, $\hat a_{\mu}$ are annihilation bosonic operators playing the role of dissipation operators accounting for quantum jumps, and $\hat H_{\rm eff}$ is an effective Hamiltonian consisting of both coherent and incoherent parts, namely,
\begin{eqnarray}
    \hat H_{\rm eff}=\hat H-\sum\limits_\mu\dfrac{i\gamma_\mu}{2}\hat a_\mu^{\dagger}\hat a_{\mu},
\end{eqnarray}
where $\hat H$ describes coherent dynamics. 

The effective NHH $\hat H_{\rm eff}$ defines the system's continuous non-unitary evolution, i.e., its dynamics between two successive quantum jumps. This continuous evolution of an open quantum system can be accessed via the postselection procedure~\cite{Minganti2020,  Naghiloo19, Abbasi2022, Chen2021, Chen2022}.

Now one notices that the Lindblad master equation in \eqref{rho} is invariant under the following affine transformations~\cite{Minganti2022}
\begin{eqnarray}\label{affine}
    \hat{K}_{\mu} &=& \hat{a}_{\mu} + \beta_\mu \hat{1}, \nonumber \\
    \hat{H}' &=& \hat{H}  - \sum_{\mu} \frac{i \gamma_\mu}{2} \left( \beta_\mu^* \hat{a}_\mu - \beta_\mu \hat{a}_\mu^\dagger \right).
\end{eqnarray} 
As a result, the effective NHH $\hat H_{\rm eff}$  changes as following
\begin{eqnarray}\label{NHH_beta}
\hat{H}_{\rm eff}(\beta) &&= \hat{H}  - \sum_\mu \frac{i \gamma_\mu}{2} \left( \beta_{\mu}^* \hat{a}_\mu - \beta_{\mu} \hat{a}_\mu^\dagger \right) \nonumber \\ 
&& \quad  -  \sum_\mu\frac{i \gamma_\mu}{2} \left( \hat{a}_\mu + \beta_{\mu} \hat{{1}} \right)^\dagger \left( \hat{a}_\mu + \beta_{\mu} \hat{{1}} \right) .
\end{eqnarray}
From \eqref{NHH_beta} it is evident that, via this invariance, the NHH can acquire an additional term, which emulates complex driving fields. 

{Physically speaking, the affine transformations in \eqref{affine} correspond to the scenario where one postselects the leaking photons via homodyne detection. 
{In a typical homodyne detection setup,} the leaking photons are mixed with an intense coherent field on a balanced beam splitter, resulting in displaced photons with a large coherent component, $\beta \gg 1$. 
In the regime where the displacement is weak, i.e., $\beta \ll 1$ {however}, this detection scheme is known as $\beta$-dyne detection~\cite{Minganti2022}. 
The $\beta$-dyne method can be implemented by mixing photons emitted from the system with an intense coherent field using an unbalanced beam splitter with low reflectivity. This $\beta$-dyne unraveling method thus allows one to introduce virtual complex driving fields in the system's effective non-Hermitian Hamiltonian~\cite{Minganti2022}.} 

This approach can be employed for the experimental realization of the NHH in \eqref{H} in an open quantum optical system operating in the single-photon regime.
Indeed, in this regime (i.e., truncating the
Fock space at the single photon), the Pauli matrices in \eqref{H} can be represented by the annihilation ($\hat a$) and creation ($\hat a^{\dagger}$) bosonic operators, namely,
\begin{eqnarray}\label{transf}
    \sigma_x^k&=&\hat a_k+\hat a_k^{\dagger}, \nonumber \\
    \sigma_y^k&=&-i(\hat a_k-\hat a_k^{\dagger}), \nonumber \\
    \sigma_z^j\sigma_z^k&=&(1-2\hat n_j)(1-2\hat n_k), \nonumber \\
\end{eqnarray}
where the photon number is defined as $\hat n_i=\hat a^{\dagger}_i\hat a_i$.

Now consider a Hermitian Hamiltonian describing a weakly driven three-mode system in a cavity with the Kerr nonlinearity in the single-photon regime
\begin{eqnarray}\label{Hbet}
    \hat H = \sum\limits_{i=1}^3\Delta_i\hat n_i+\sum\limits_{j<k}U_{jk}\hat n_j\hat n_k+\sum\limits_{i=1}^3(\eta_i^*\hat a_i+\eta_i\hat a_i^{\dagger}).\nonumber \\
\end{eqnarray}
 We further assume  that the cavity is lossy; that is, each mode $\hat a_i$ is characterized by the photon decoherence rate $\gamma_i$. Then, by monitoring the leaking photons from the cavity via the $\beta$-dyne detection, one can induce the continuous non-unitary evolution in the system governed by the  effective NHH which can be identified with that in the \eqref{H}. 
Namely, by combining Eqs.~(\ref{NHH_beta})--(\ref{Hbet}), and by introducing the following relations
\begin{eqnarray}\label{param2}
    \Delta_i&=&-2\sum\limits_{j<k} J_{jk}-\dfrac{i\gamma_i}{2}, \nonumber \\
    U_{jk}&=&4J_{jk}, \nonumber \\
    \gamma_k\beta_k&=&2\left[{\rm Im}(B_x^k)+i{\rm Im}(B_y^k)\right], \nonumber \\
    \eta_k&=&(B_x^k)^*+i(B_y^k)^*,
\end{eqnarray}
 {the resulting effective $\beta$-dyne NHH can be exactly mapped to the NHH in \eqref{H}, provided that $\gamma_i=\gamma$ for $\forall i$, such that the modes' decay rate $\gamma/2$ can be gauged away. However, the latter condition can be eased when the difference $|\gamma_i-\gamma_j|\neq0$, $i\neq j$, but it is still small enough ensuring that the protocol remains stable. We also refer the interested reader to Ref.~\cite{Minganti2022}, which provides a detailed discussion on the physics of the $\beta$-dyne detection scheme.}

\section{Effects of perturbations on dynamical state permutations in a three-qubit system}\label{Perturbation}

Here, we briefly examine the impact of perturbations in the three-qubit NHH on the state permutations discussed in Sec.~\ref{Group}. Generally, arbitrary perturbations to the NHH in \eqref{H} lead to a complex spectrum with a non-zero imaginary part, which plays a critical role in triggering NATs in the system dynamics~\cite{Doppler_2016}. For sufficiently large perturbations, the permutation symmetry of the states is broken, resulting in chiral state dynamics. However, when perturbations remain small, the symmetry in state flips can be preserved.

\begin{figure*}[t!]
    \includegraphics[width=0.99\textwidth]{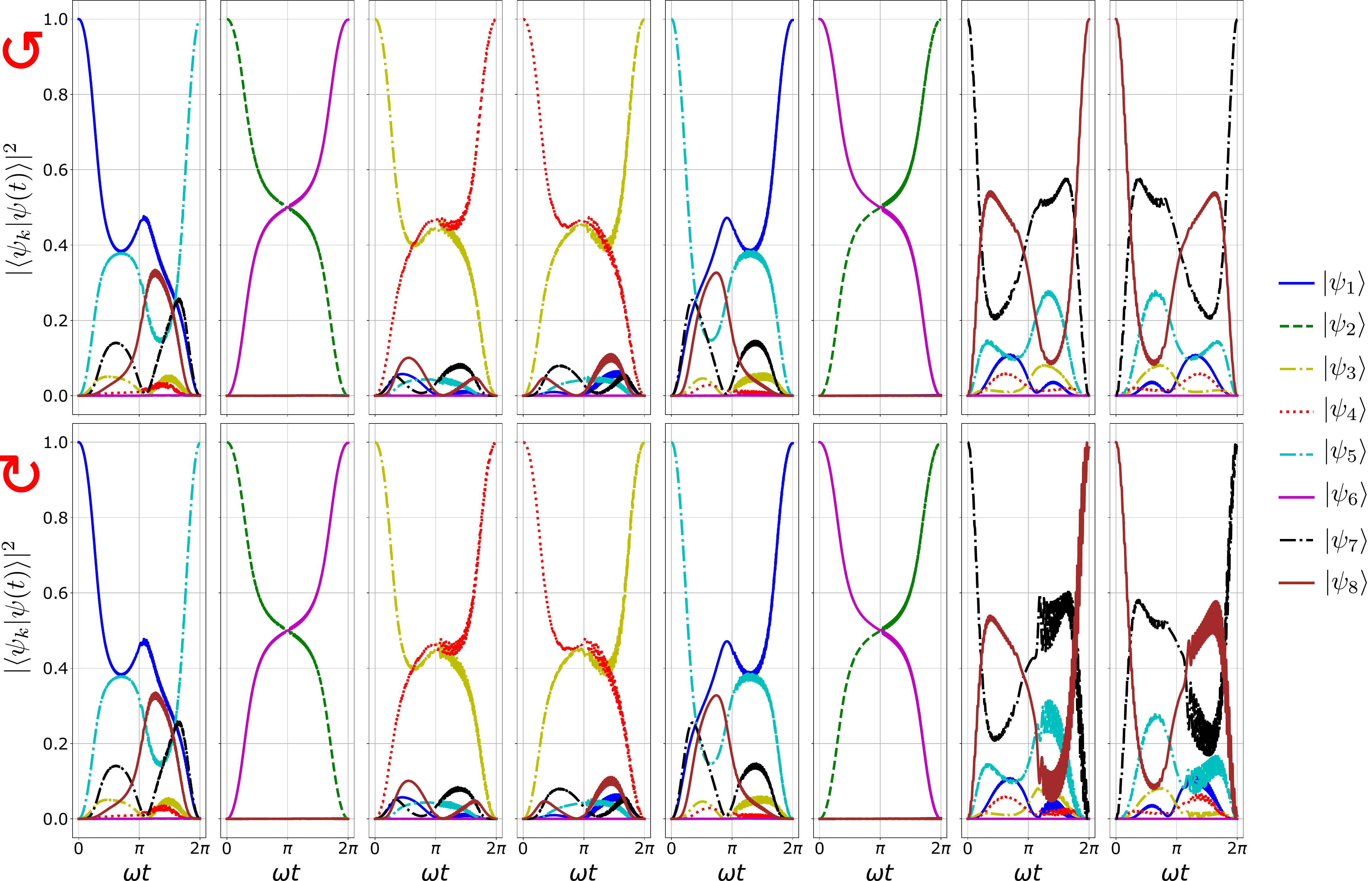}
    \caption{
    State transfer fidelity for the perturbed three-qubit NHH, according to \eqref{H} and \eqref{Hpert} for a perturbation $\epsilon=0.001$. This figure corresponds to the unperturbed case shown in \figref{fig3}(d). The upper (lower) row of panels demonstrates the scenario when the dynamical winding is performed in the counterclockwise $+\omega$ (clockwise $-\omega$) direction. The time period is chosen as $T=500$ [arb. units], the remaining system parameters are the same as in \figref{fig3}.
    }
    \label{S6}
\end{figure*}

\begin{figure*}[t!]
    \includegraphics[width=0.99\textwidth]{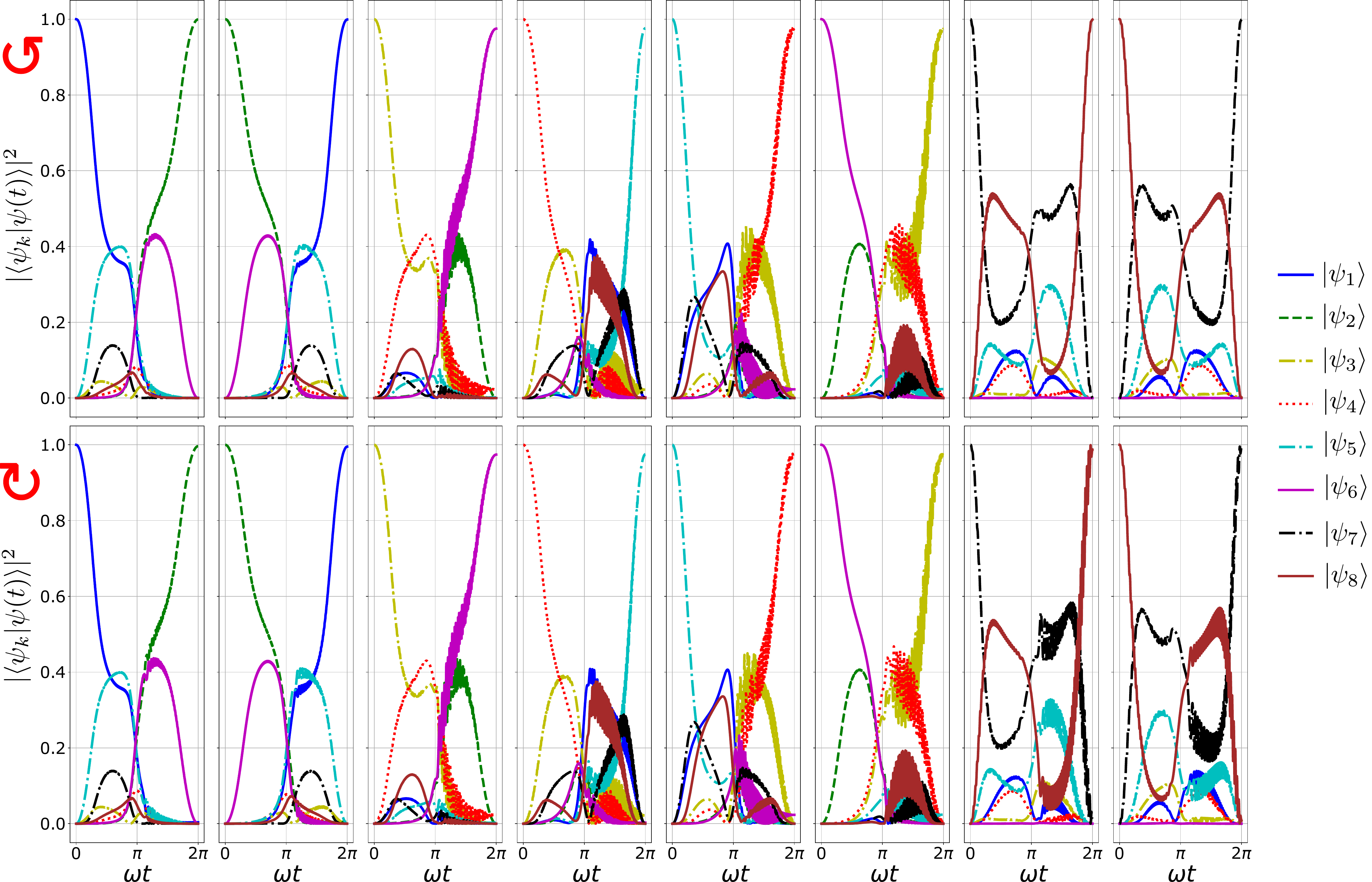}
    \caption{
    State transfer fidelity for the perturbed three-qubit NHH, according to \eqref{H} and \eqref{Hpert} for a perturbation $\epsilon=0.001$. This figure corresponds to the unperturbed case shown in \figref{fig4}(d). The upper (lower) row of the panels shows the scenario when the dynamical winding is performed in the counterclockwise $+\omega$ (clockwise $-\omega$) direction. The time period is chosen as $T=500$ [arb. units], , the remaining system parameters are the same as in \figref{fig4}.
    }
    \label{S7}
\end{figure*}

Without loss of generality, we assume that a three-qubit NHH $\hat H$ in \eqref{H} is perturbed as follows 
\begin{eqnarray}\label{Hpert}
    H\to H+\epsilon \sigma_z^1. 
\end{eqnarray}
That is, there is a non-zero magnetic field perturbation applied to the first qubit along the $z$-axis. The effects of this perturbation on the state transfer protocol are shown in \figref{S6} and \figref{S7}. Figure~\ref{S6} (\figref{S7}) corresponds to the unperturbed case shown in \figref{fig3} (\figref{fig4}). As seen in both figures, the state evolution under perturbation starts to differ based on the winding direction, with this difference growing as the disorder increases, eventually leading to state evolution instabilities and NATs (not shown).

However, as demonstrated in Figs.~\ref{S6} and \ref{S7}, for small levels of disorder ($\epsilon \approx 0.001$), the desired state permutations remain feasible. By appropriately tuning the system parameters, the state transfer fidelity can still be kept at high rates.


\section{Numerical simulations of state dynamics and solution stiffness}\label{AG}
\begin{figure*}[t!]
    \includegraphics[width=0.99\textwidth]{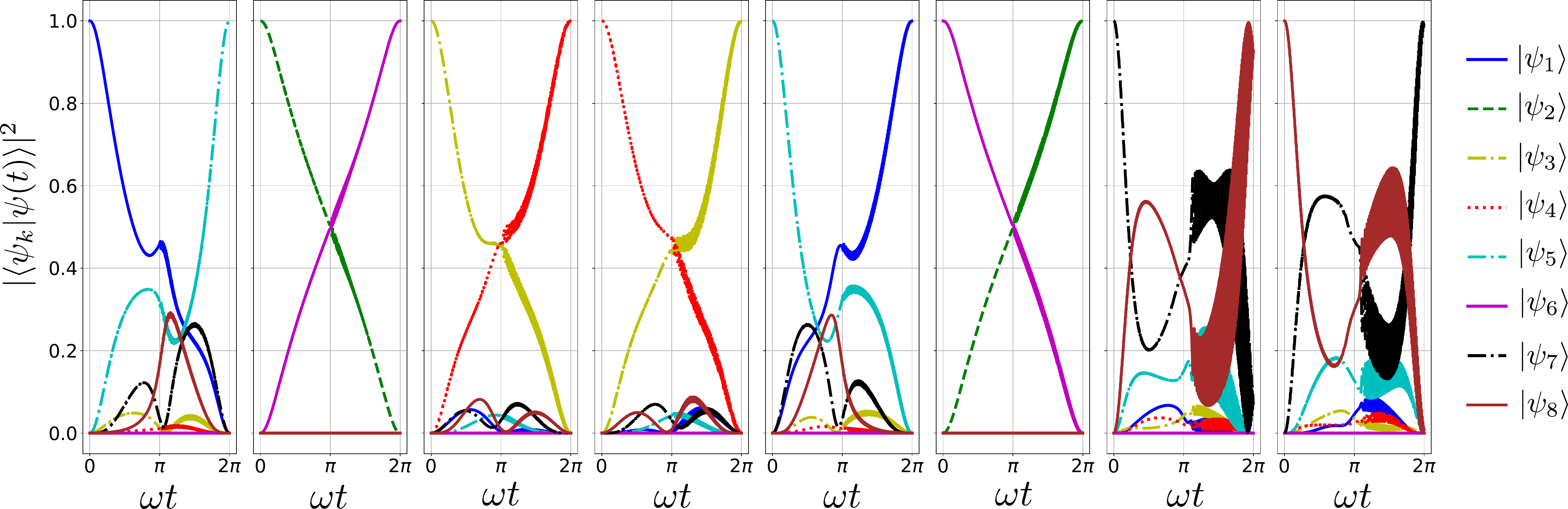}
    \caption{Fidelity $|\langle \psi_k|\psi(t)\rangle|^2$ between the evolving state $|\psi(t)\rangle$ and the initial eigenstate $|\psi_k\rangle$, $k=1,\dots,8$, of the NHH. This figure corresponds to the \figref{fig4} in the main text, with only difference that the chosen trajectory loop is a circle with the radius $r=1$ in \eqref{param}. The rest system parameters are the same. For trajectory the dynamics for the states $|\psi_7\rangle$ and $|\psi_8\rangle$ exhibits rapid oscillations in the time interval $T/2<t<T$, due to stiffness of the solution (see the text in the Appendix~\ref{AG} for more details).
    }
    \label{S8}
\end{figure*}

In this section, we address the issue of solution stiffness in the simulated state dynamics. Specifically, we observe that when calculating the fidelity for the state-switching between $\psi_7$ and $\psi_8$, as shown in~\figref{fig4}, the fidelity might exhibit rapid oscillations for certain encircling loops. For instance, in \figref{S8}, we plot the state fidelity for the same scenario as in \figref{fig4}, but with the encircling loop along a circle of radius $r=1$. One can see that during the transition $\psi_7 \to \psi_8$, after crossing the branch cut, the fidelity undergoes rapid oscillations with a non-negligible amplitude, which gradually vanishes as the system approaches the final state $\psi_8$.

This stiffness issue arises because the energy sheets $E_7$ and $E_8$ are very smooth and lie very close to each other when the evolving states transition between these levels near the branch cut. Consequently, the simulated state dynamics may exhibit rapid variations, which can be suppressed by taking very small integration steps when solving the Schrödinger equation, or by varying the shape of the enclosing loop such that the stiffness can be mitigated (as shown in the main text).
The interested reader may refer to Ref.~\cite{Aiken1985}, where the problem of stiff equations is discussed in greater detail.

\begin{figure*}[t!]
    \includegraphics[width=0.99\textwidth]{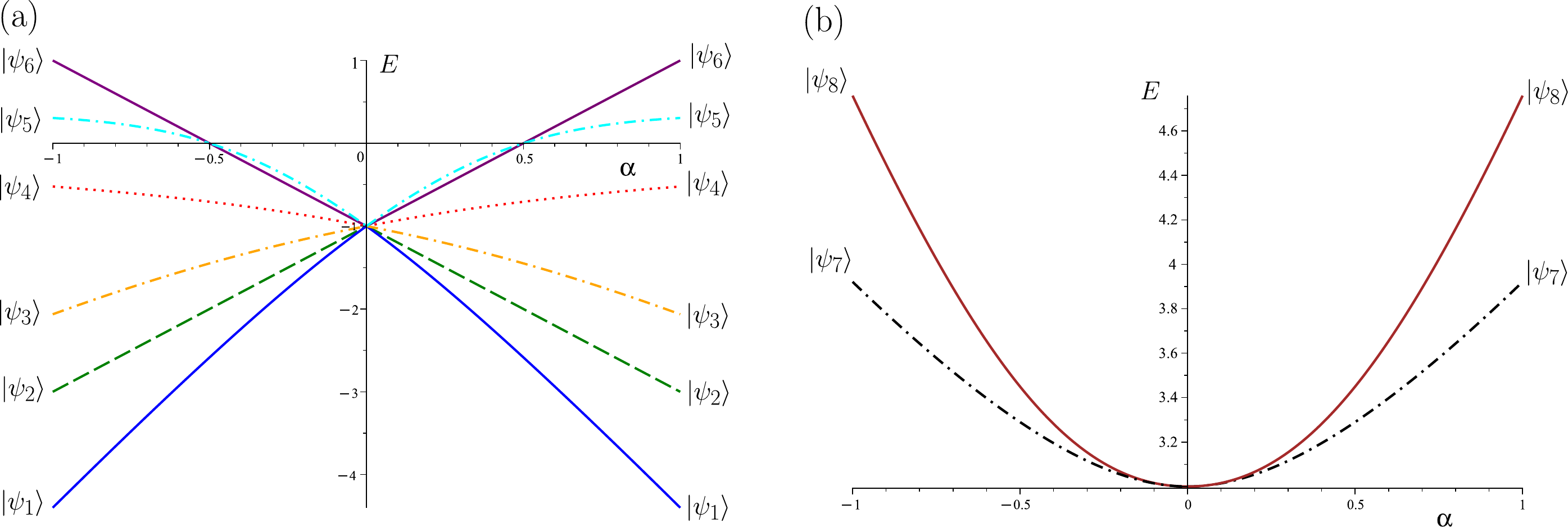}
    \caption{Energy levels of the three-qubit system, studied in Sec.~\ref{IV} in the main text, as a function of the parameter $\alpha$. (a) The energy levels, corresponding to the first six eigenstates $|\psi_{i}\rangle$, $i=1,\dots,6$, are linear in the vicinity of the branch cuts $\alpha=0$ and have large slopes with respect to zero, ensuring the solution stability of the state dynamics. (b) The energy levels of the eigenstates $|\psi_{7,8}\rangle$ have mostly zero slope near the branch cut and lie close to each other. This can result in rapid oscillations in the state dynamics for some chosen trajectories, as shown in \figref{S8} (see also the text in the Appendix~\ref{AG} for more details).
    }
    \label{S9}
\end{figure*}

\begin{figure*}[h!]
    \includegraphics[width=0.8\textwidth]{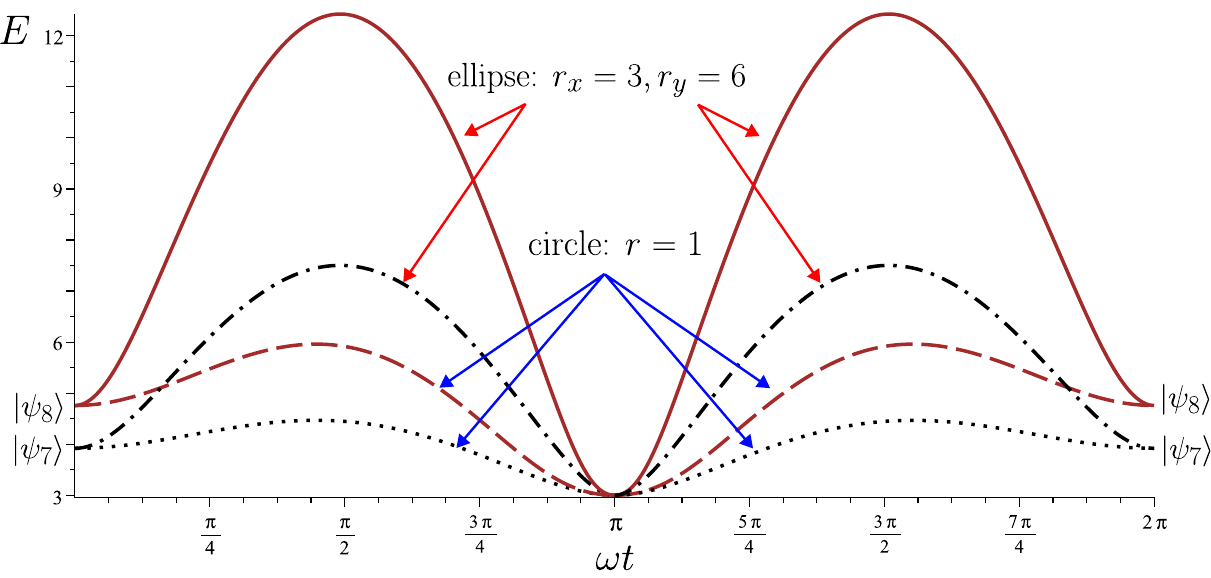}
    \caption{Energy levels corresponding to the eigenstates $|\psi_7\rangle$ (black curves) and $|\psi_8\rangle$ (brown curves) along the two different loops around the EP in the system parameter space, according to \eqref{param}. The energy levels along the elliptic loop, shown as brown solid and black dash-dotted curves, correspond to the scenario in \figref{fig4}. The energy levels along the circle with the radius $r=1$, shown as brown dotted and black dotted curves, correspond to the scenario in \figref{S8}. For the circle trajectory, the energy levels are very smooth and lie close to each other which can lead to rapid oscillations in the state dynamics (as shown in \figref{S8}). These oscillations can be mitigated by choosing an appropriate trajectory, e.g., an elliptic loop, where the relative slope between the two energy levels becomes larger.
    }
    \label{S10}
\end{figure*}

In \figref{S9}, we plot the energy levels as a function of $\alpha$, corresponding to the scenario shown in \figref{fig4}. As seen in panel (a), for the first six eigenstates, the energy dependence around the branch cut $\alpha = 0$ is linear. This linear dependence ensures stable dynamics, as confirmed in \figref{S8}. However, for the states $\psi_7$ and $\psi_8$, the energy curves are significantly smoother near $\alpha=0$, as shown in \figref{S9}(b). Indeed, by expanding the energies $E_{7,8}$ in a power series in $\alpha$, they take a general form $E_{7,8}=u_2^{7,8}\alpha^2+u_3^{7,8}\alpha^3+\dots$, where $u^{7,8}_i\in{\mathbb R}$.  This smoothness, i.e., the nearly zero slope of the energy curve, and proximity of the energy levels can thus lead to rapid oscillations in the solutions for the state dynamics. However, as we have demonstrated in the main text, this numerical difficulty can be overcome by choosing, e.g., elliptic loops in the state trajectories, characterized by the large major semi-axis along the $Oy$ axis, which increases the relative slope between the energy curves near the EP and, therefore, can restore the solution stability (see also \figref{S10} and its comparison with Figs.~\ref{fig4} and \ref{S8}).


\clearpage
%

\end{document}